\documentclass[transmag]{IEEEtran}
\usepackage{latexsym}

\IEEEoverridecommandlockouts
\usepackage{amsmath,graphicx, amssymb, amsfonts}
\usepackage{booktabs}

\usepackage[bookmarks=false]{hyperref}
\usepackage[T1]{fontenc}
\usepackage[font=footnotesize]{caption}
\usepackage{color,soul}
\usepackage{float}
\usepackage{url}

\definecolor{gray}{rgb}{1,0.6,0.5}
\setstcolor{gray}
\sethlcolor{gray}
\usepackage{multirow}
\usepackage{siunitx}
\usepackage{enumitem}
\usepackage{cite}
\usepackage{subcaption}
\usepackage{dsfont}
\usepackage[thinc]{esdiff}
\usepackage[pdftex,dvipsnames]{xcolor} 
\usepackage[colorinlistoftodos,prependcaption,textsize=small]{todonotes}

\usepackage{url}

\usepackage{breakurl}

\usepackage{diagbox}

\usepackage[linesnumbered,ruled,vlined]{algorithm2e}
\usepackage{algpseudocode}

\usepackage{tikz}
\usetikzlibrary{matrix,chains,positioning,decorations.pathreplacing,arrows,spy}
\tikzset{every picture/.style={font issue=\footnotesize},
	font issue/.style={execute at begin picture={#1\selectfont}}
}
\usetikzlibrary{shapes.geometric}

\newlength{\Oldarrayrulewidth}

\def\BibTeX{{\rm B\kern-.05em{\sc i\kern-.025em b}\kern-.08em T\kern-.1667em\lower.7ex\hbox{E}\kern-.125emX}}
\begin{document}
	
	\title{Efficient Bitrate Ladder Construction for Content-Optimized\\ Adaptive Video Streaming}
	
	\author{\IEEEauthorblockN{Angeliki~V.~Katsenou$^*$, Joel Sole$^\dagger$ and David~R.~Bull$^*$}
		\thanks{Submitted on December 15, 2020. The work presented was supported by the Leverhulme Early Career Fellowship (ECF-2017-413) and by Netflix Inc.}
		\thanks{A. Katsenou and D. R. Bull, are with University of Bristol, Bristol, BS1 5DD, UK.  (e-mail: \{angeliki.katsenou, dave.bull\}@bristol.ac.uk)}
		\thanks{J. Sole is with 
			Netflix, Inc, Los Gatos, California, USA (e-mail: jsole@netflix.com).}}

	\IEEEtitleabstractindextext{\begin{abstract} One of the challenges faced by many video providers  is the heterogeneity of network specifications, user requirements, and content compression performance. The universal solution of a fixed bitrate ladder is inadequate in ensuring a high quality of user experience without re-buffering or introducing annoying compression artifacts. However, a content-tailored solution,  based on extensively encoding across all resolutions and over a wide quality range is highly expensive in terms of computational, financial, and energy costs. Inspired by this, we propose an approach that exploits machine learning to predict a content-optimized bitrate ladder. The method extracts spatio-temporal features from the uncompressed content, trains machine-learning models to predict the Pareto front parameters and, based on that, builds the ladder within a defined bitrate range. The method has the benefit of significantly reducing the number of encodes required per sequence. The presented results, based on 100 HEVC-encoded sequences, demonstrate a reduction in the number of encodes required when compared to an exhaustive search and an interpolation-based method, by 89.06\% and 61.46\%, respectively, at the cost of an average Bj{\o}ntegaard Delta Rate difference of 1.78\% compared to the exhaustive approach. Finally, a hybrid method is introduced that selects either the proposed or the interpolation-based method depending on the sequence features. This results in an overall 83.83\% reduction of required encodings at the cost of an average Bj{\o}ntegaard Delta Rate difference of 1.26\%. \end{abstract}
		
	}
	
	\maketitle
	
	\section{Introduction}
	\label{sec:Intro}
	
	\IEEEPARstart{I}{n} recent reports on internet traffic volumes~\cite{Cisco}, the share occupied by video data is predicted to reach 80\% by 2023 with anticipation of further rises subsequently. Due to the recent pandemic and the associated major shift towards remote working (work from home schemes, online education, etc)~\cite{CapacityArticle2020}, this figure is now expected to be reached even sooner. Although mobile users are exchanging more and more of the content they generate, the major share of the video networking index relates to on-demand video services, such as those provided by Netflix, Hulu, Amazon Prime, and others. All video service providers invest a significant amount of resource into optimizing video compression parameters prior to transmission. This enables them to increase user satisfaction - meeting varying end-user constraints while maintaining the highest possible level of delivered video quality.
	
	The display quality of the delivered content may vary from device to device, and may be affected by a variety of factors such as location, terminal equipment type and available bandwidth. For example, a given mobile phone is likely to receive a different encoded version of the same source video  on a 5G network than it would on a 4G network. These encodes may vary in terms of both compression ratio and spatial resolution. It also means that an end-user's device might receive content compressed at lower resolutions that is then upscaled to a device's native resolution prior to display.
	
	Many of the video service providers adopt HTTP Adaptive Streaming (HAS) through
	Dynamic Adaptive Streaming over HTTP (DASH), which is the standard solution introduced by MPEG~\cite{MPEG-Dash}. In DASH, videos are encoded with a different set of parameters (resolution, quantization parameter, etc.) to allow for the adaptation of the delivered video content to the heterogeneous network conditions (available bandwidth, device characteristics, etc) so as to ensure high quality of experience. The traditional approach has been to construct (at the server side) a bitrate ladder, which constitutes a set of bitrate-resolution pairs. This type of bitrate ladder is often referred to as ``one-size-fits-all''. In early examples, two bitrate points were used for 1080p: 4300 kbps and 5800 kbps regardless of title~\cite{AaronNtflx2015}. In later advances, differentiation  was introduced based on the genre of the content, e.g.~\cite{LedererMMSys2012}. For example, higher bitrates were used for sports content with rapid motion and fast scene changes. These solutions, however, ignored the dependency of the video compression performance on specific content  characteristics,  resulting in noticeable blocking and other visual artifacts in some cases and thus, in a degraded viewing experience.
	
	Recently, content-customised solutions have been developed and adopted by industry, such as those used by Netflix~\cite{AaronNtflx2015,NetflixDOICIP2016, NtflxBlog2020}. The key task here  is to invest in pre-processing where each video title is split into shorter clips or chunks, usually associated with shots. Each short video chunk is encoded using optimized parameters, i.e. resolution, quantization level, intra-period, etc, with the aim of building the Pareto Front (PF) across all Rate-Quality (RQ) curves. After that, a set of target bitrates is used to find the best encoded bitstreams. Given the extensive parameter space (compression levels, spatial and temporal resolution, codec type etc.) and taking into account the fact that this process must be repeated for each video chunk, the amount of computation needed is massive. As a consequence, the industry heavily relies on cloud computing services for pre-processing, and this naturally comes with a high cost in financial, time and environmental terms.
	
	Considering the above, in this paper we propose a content-gnostic method of estimating a close-to-optimal bitrate ladder for adaptive video streaming. The proposed method is based on extracting low-level content features from the uncompressed videos at their native spatial resolution and on training machine-learning models to predict the PF parameters of the rate-distortion curves across different resolutions. Based on the estimated PF parameters, a set of equations that model the quantization parameters to the bitrate is defined. With this set of equations and taking into account the available bitrate range, a suitable bitrate ladder is constructed per video sequence. A clear benefit of the proposed method compared to previous approaches is that it significantly reduces the amount of computation required. Furthermore, since the bitrate ladder is constructed using the RQ PF, the number of steps on the ladder is not fixed and might be reduced for certain videos compared to other bitrate ladder solutions. This further reduces the storage requirements for the resulting encodes.

	\subsection{Contribution}
	In our previous work~\cite{KatsenouPCS2019}, we proposed a content-gnostic method that predicts the cross-over points of RQ curves across three spatial resolutions (2160p, 1080p, and 720p) for each video chunk based on Peak Signal-to-Noise Ratio (PSNR). Our method had the unique characteristic of predicting the Quantization Parameters (QPs) associated with the curve intersection point without requiring any encodings. The encodings were performed only to estimate the bitrates corresponding to the switching of resolutions.
	
	In this paper, we build upon this previous work and propose an extension to it. The new framework extends beyond the prediction of the RQ intersection points across spatial resolutions by introducing the following:
	\begin{itemize}
		\item A new content-driven process to predict the bitrate ladder is introduced that takes into account both bitrate and quality constraints. This method models the relationship between rate and quantization parameters across resolutions and utilises this for the estimation of the bitrate ladder.
		\item Further to the feature-based method, a hybrid methodology is proposed. This method selects either the content driven method or an interpolation-based method for an input sequence as a solution that can balance the accuracy of prediction to the relative complexity trade-off.
		\item The test case presented is based on an extended set of resolutions from 2160p down to 540p, i.e. 3840$\times$2160, 1920$\times$1080, 1280$\times$720, 960$\times$540, and on a bitrate range typical for video streaming at these resolutions (from 150kbps to 25Mbps).
	\end{itemize}

	\subsection{Paper Organization}
	The rest of this paper has the following structure. Firstly, Section~\ref{sec: RelatedWork} outlines state-of-the-art research and industrial technologies. In Section~\ref{sec: Outline}, the proposed framework is introduced. The dataset employed together with its low-level features are described in Section~\ref{sec: ProposedMethod}. In the same section, the modules that contribute to the construction of the bitrate ladder are detailed. Next, in Section~\ref{sec: ConvHullPred}, the extracted spatio-temporal features and the machine learning techniques used for the prediction of the bitrate ladder are reported. Moreover, this section presents results on the predicted ladder and discusses the methods' relative complexity. Finally, conclusions and future work are summarised in Section~\ref{sec: concl}.
	
	\begin{figure*}[!ht]
		\centering
		\includegraphics[width=\linewidth]{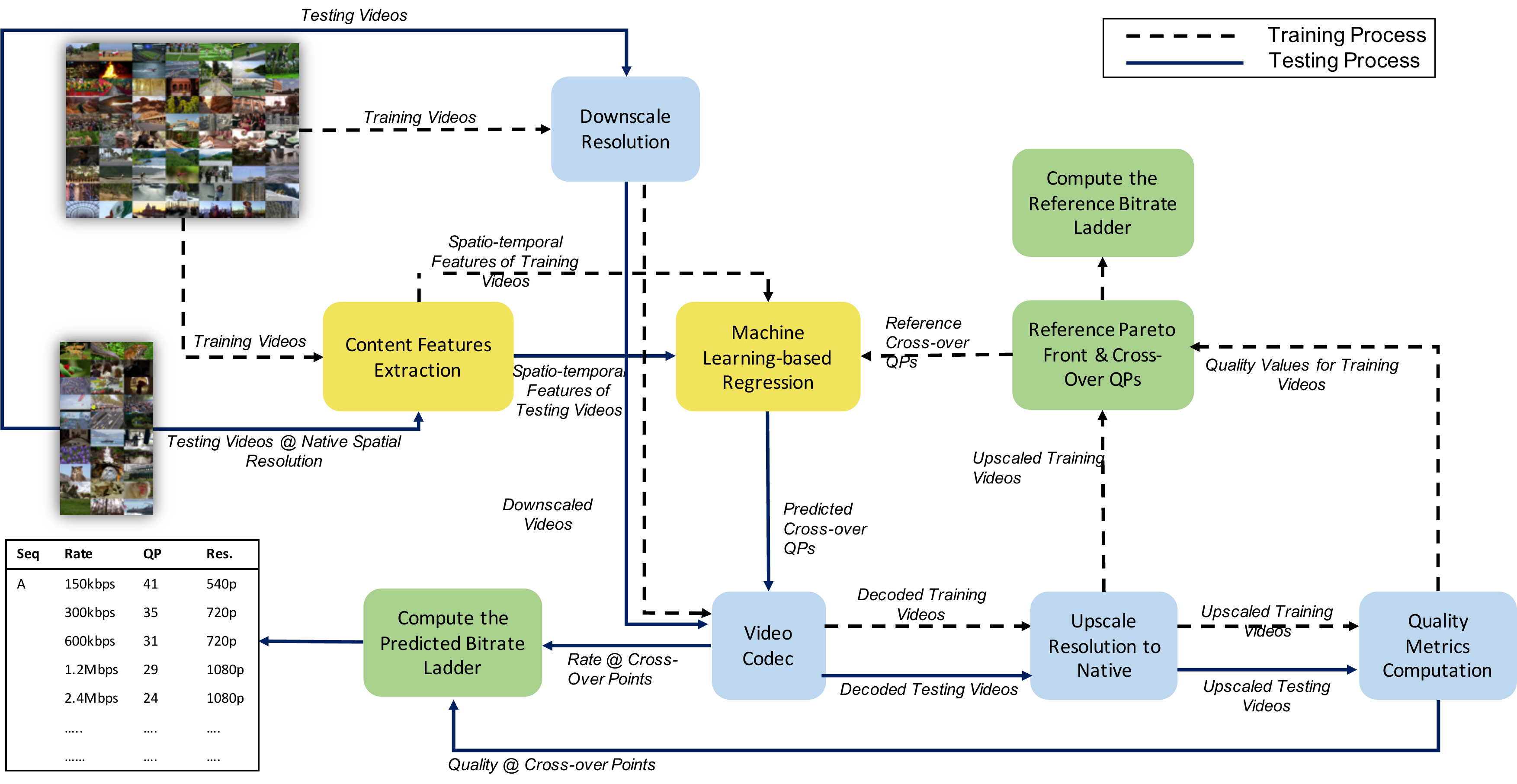}
		\caption{Diagrammatic overview of the proposed framework. Blue blocks indicate off-the-shelf technologies/tools, yellow the methodologies employed in our previous work~\cite{KatsenouPCS2019}, and green blocks the new processes introduced in this paper.}
		\label{fig: method}
	\end{figure*}
	
	\section{Related Work}
	\label{sec: RelatedWork}
	As explained in the Introduction, the traditional approach employed builds a fixed ``bitrate ladder'' that yields recommended spatial resolution for the available bitrate. The bitrate ladder is either content-agnostic, with fixed bitrate ranges allocated per resolution, e.g.~\cite{AppleStaticLadder}, or employs a limited number of bitrate ladders based on genre, e.g.~\cite{LedererMMSys2012}. 
	
	More advanced methods that move beyond the fixed ``bitrate ladder'' approach have been proposed recently by both academic and industry stakeholders~\cite{NetflixDOICIP2016,DynOptimiser,NtflxBlog2020,KokaramICIP2018,TimmererICIP2018,Toni2015, BitmovinBitrateLadder, MUX, Cambria, Ozer2018, Reznik2018, Giladi2018, TimmererICME2020}. All of these solutions rely on encoding statistics collected either by performing a massive number of encodings or by using a selective number of encodes to predict the bitrate ladders using pre-trained machine learning models.
	
	Most of these recently introduced methods were based on a per-title optimization framework.
	Netflix~\cite{AaronNtflx2015,NetflixDOICIP2016,DynOptimiser} have reported that they obtain the RQ curves per title at different resolutions and at different bitrates by running several trial encodings at different quantization levels. Then, this information is used to construct the Pareto-optimal front (often referred to in the adaptive streaming literature as a convex hull) of the RQ curves using both scaled PSNR~\cite{NetflixDOICIP2016} and scaled VMAF in~\cite{DynOptimiser} and hence to obtain the optimal parameters for a defined bitrate range. 
	An alternative approach, presented in~\cite{KokaramICIP2018}, uses measurements on the actual usage of millions of video clips to create probability distributions of available bandwidth and viewport sizes. These probability distributions feed an optimization process that ensures video quality preservation while reducing the required bitrate compared with existing techniques. Other per-title-encoding approaches have been developed by Bitmovin~\cite{BitmovinBitrateLadder}, MUX~\cite{MUX}, CAMBRIA~\cite{Cambria} and others~\cite{Ozer2018}. The Bitmovin~\cite{BitmovinBitrateLadder} and CAMBRIA~\cite{Cambria} solutions include computation of the encoding complexity. According to the former~\cite{BitmovinBitrateLadder}, a complexity analysis is performed on each incoming video, and a variety of measurements are processed by a machine-trained model to adjust the encoding profile to match the content. The CAMBRIA solution estimates the encoding complexity by running a fast constant rate encoding~\cite{Cambria}. 
	MUX~\cite{MUX} introduced a deep-learning based approach that takes, as input, the vectorized video frames and predicts the bitrate ladder. The aforementioned approaches are compared in~\cite{Ozer2018}. A further approach that uses trial encodes to collect coding statistics at low resolutions and utilizes them within a probabilistic framework to speed up the encoding decisions at higher resolutions is presented in~\cite{Giladi2018}.
	
	Recent work has presented a per scene optimization method which aims to either maximize the quality or minimize the bitrate of each encoded representation in video on demand HAS scenarios~\cite{TimmererICME2020}. The method relies on building a quantized convex hull by encoding the sequences across a set of spatial resolutions. Furthermore, Netflix~\cite{NtflxBlog2020} has updated its dynamic optimization method by building a bitrate ladder after complexity analysis and by further tuning of pre-defined encoding parameters. Another interesting approach that takes into account both quality constraints and bitrate network statistics was proposed by Brightcove~\cite{Reznik2018,ReznikICME2019}. The quality metric used in this case was the Structural Similarity Index Measure (SSIM) and bitrate constraints were based on probabilistic models. Another recent approach, the iSize solution~\cite{iSize}, uses pre-encodes within a deep learning framework to decide on the optimal set of encoding parameters and resolution at a block level. 
	
	While all of the above solutions are significant and have contributed in the enhancement of video services, it is not possible to make direct detailed comparisons. Firstly, they are proprietary, meaning that they are designed to satisfy differing business requirements and that their full details may not be public~\cite{TimmererSurvey}. Moreover, some of the methods are using different metrics in their bitrate adaptation process that are not always shared. One area of improvement for these methods is to reduce their complexity, as most of the aforementioned solutions rely on large numbers of encodes (in many cases massive). Hence the computational, energy and financial costs are high, since cloud encoding services are usually employed~\cite{OzerPricing}. 
	
	In the above context, our aim here is to build a methodology that can predict a close to optimal content-gnostic bitrate ladder at a reduced computational cost compared to traditional methods. The proposed methodology is outlined in the following section.

	\section{Outline of the Proposed Framework}
	\label{sec: Outline}
	
	The proposed framework, as shown in Fig.~\ref{fig: method}, is structured in two processes that use common functionalities; the training and testing processes. Different line patterns are used to denote the information flow for these.
	During the training process, we first downscale the uncompressed sequences to create different spatial resolution versions. We also extract low-level content features from the native resolution sequences. Then, we encode the native and downscaled sequences for a wide range of QPs. After decoding, we rescale all versions in order to compute quality metrics at the native resolution and construct the reference Pareto Front (PF) of the sequence. Simultaneously, we record the intersection points of the RQ curves across resolutions, that is, the bitrate, quality, QP, and resolution. The QP values  at the intersection points between each resolution are henceforth called cross-over QPs. The QP values represent the independent variable in the encoding process (rate and quality are the dependent variables), and assume discrete values. Thus, our first step is to predict these QP values as a basis for bitrate prediction. The cross-over QPs resolutions of the reference PF represent the ground truth for our predictions.
	
	Using the extracted spatio-temporal features, the ground truth cross-over points and the associated PFs, we train supervised machine learning models to perform regression and predict the cross-over points. Using the reference PF, we can construct the reference bitrate ladder, as follows: firstly we reduce the bitrate range within practical limits for streaming; secondly, this trimmed Pareto surface is subsampled across the quality and bitrate dimensions, as detailed in Section~\ref{ssec: BuildLadder}.
	
	The testing process is similar, but simpler than the training process. We first extract the spatio-temporal features that were selected during the training process from the uncompressed test sequences at the native resolution. Then, we use the trained models to predict the cross-over QPs. These QPs help to determine the bitrates at which the resolution switches to the next one on the PF. Thus, we perform a small number of encodes across resolutions at the cross-over QPs. The next step involves defining the parameters of the set of equations that relate the quantization parameter to the bitrate. These equations will indicate the resolution and QP at the target bitrate ladder rungs. The final step involves encoding the sequence at the predicted resolution and QPs for each ladder rung so as to derive the predicted content-aware bitrate ladder.
	
	While the proposed methodology has been implemented and demonstrated using the High Efficiency Video Codec (HEVC) codec~\cite{HEVCpaper}, it is extendable to any video codec once properly trained. Furthermore, in this paper, we selected PSNR as the basis for constructing the bitrate ladder. PSNR remains the most commonly used quality metric despite the fact that other quality metrics have been shown to offer a better correlation with perceived quality. Nevertheless, the method is adaptable to other quality metrics.
	
	\section{Constructing the Reference Bitrate Ladder}
	\label{sec: ProposedMethod}
	This section focuses on the exploration of the RQ space across resolutions, the definition and modelling of the reference Pareto surface, and the construction of the sequence-specific reference bitrate ladder.
	\subsection{Description of the Dataset}
	\label{ssec: dataset}
\begin{figure}[!t]
		\centering
		\begin{minipage}[b]{\linewidth}
			\centering
			\includegraphics[width=\linewidth, trim={0 0 0 0},clip]{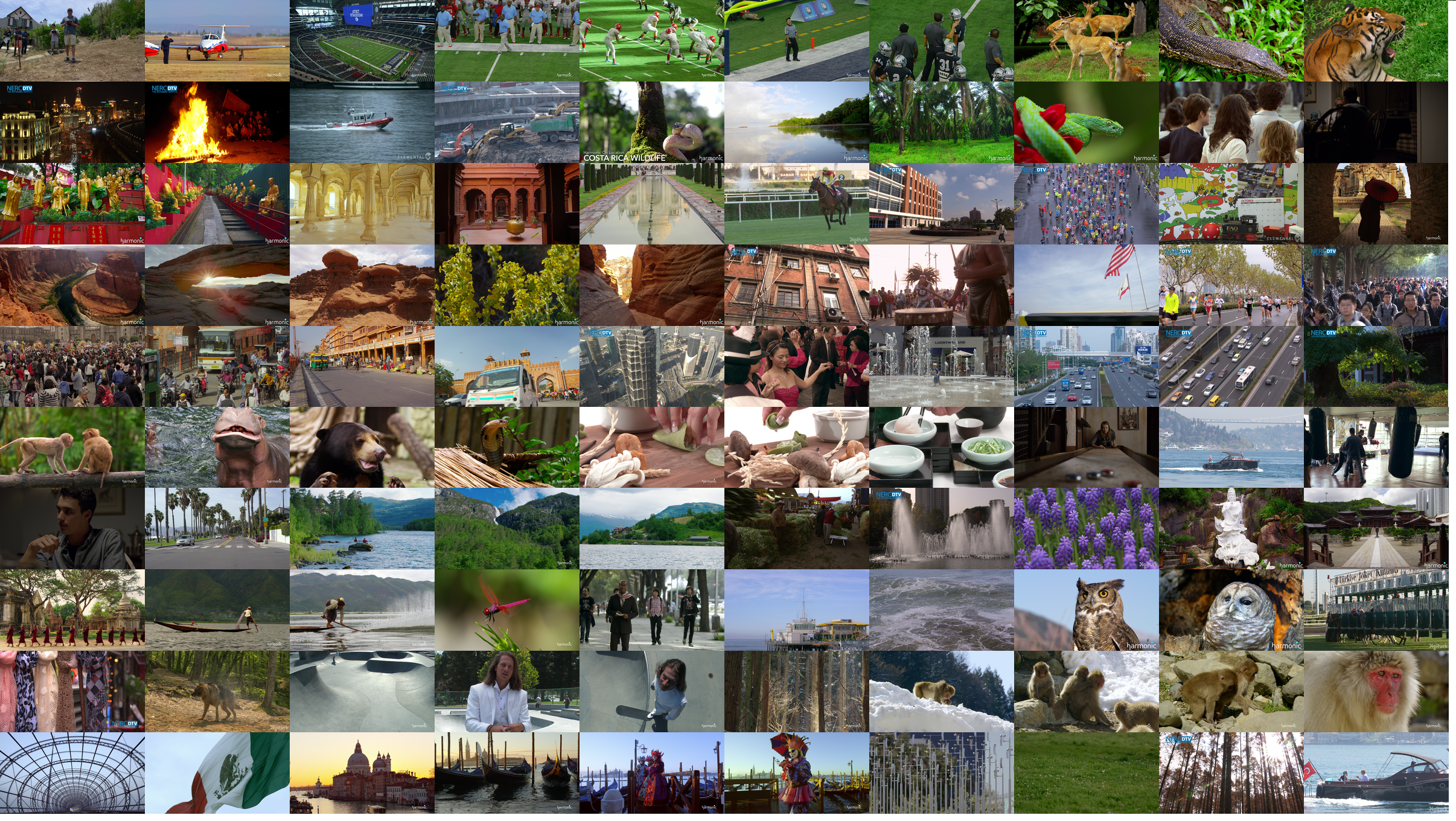}
		\end{minipage}
		\caption{Sample frames of the considered dataset~\cite{AfonsoSPIE2018}.} 
		\label{fig:Sample100}
	\end{figure}

	For any content-driven video processing framework, it is essential to have a large video dataset that covers a variety of scenes. Therefore, we employed a dataset of 100 publicly available UHD video sequences from different sources: Netflix Chimera~\cite{ioannis2015Netflix}, Ultra Video Group \cite{tampereTest}, Harmonic \cite{harmonicFootage}, SJTU \cite{song2013sjtu} and AWS Elemental \cite{AWSElemental}. The same dataset was also used as a training dataset in~\cite{AfonsoSPIE2018}. Example frames from the dataset are depicted in Fig.~\ref{fig:Sample100}. Many of the sequences have a native resolution of 4096$\times$2160 and were spatially cropped to 3840$\times$2160 and converted to a 4:2:0 chroma subsampled format (if originally otherwise). Each sequence contains a single scene (no scene cuts) and the scenes are representative of different objects/regions of interest, camera motions, colours, and spatial activity. The majority of the test sequences have a frame rate of 60 fps\footnote{Two of the sequences were temporally downsampled from 120 to 60 fps in order to match the majority frame rate of 60 fps.} and a bit depth of 10 bits per sample. Finally, the sequences were temporally cropped to 64 frames.

\begin{figure}[!ht]
		\centering
		\begin{minipage}[b]{.49\linewidth}
			\centering
			\includegraphics[width=\linewidth]{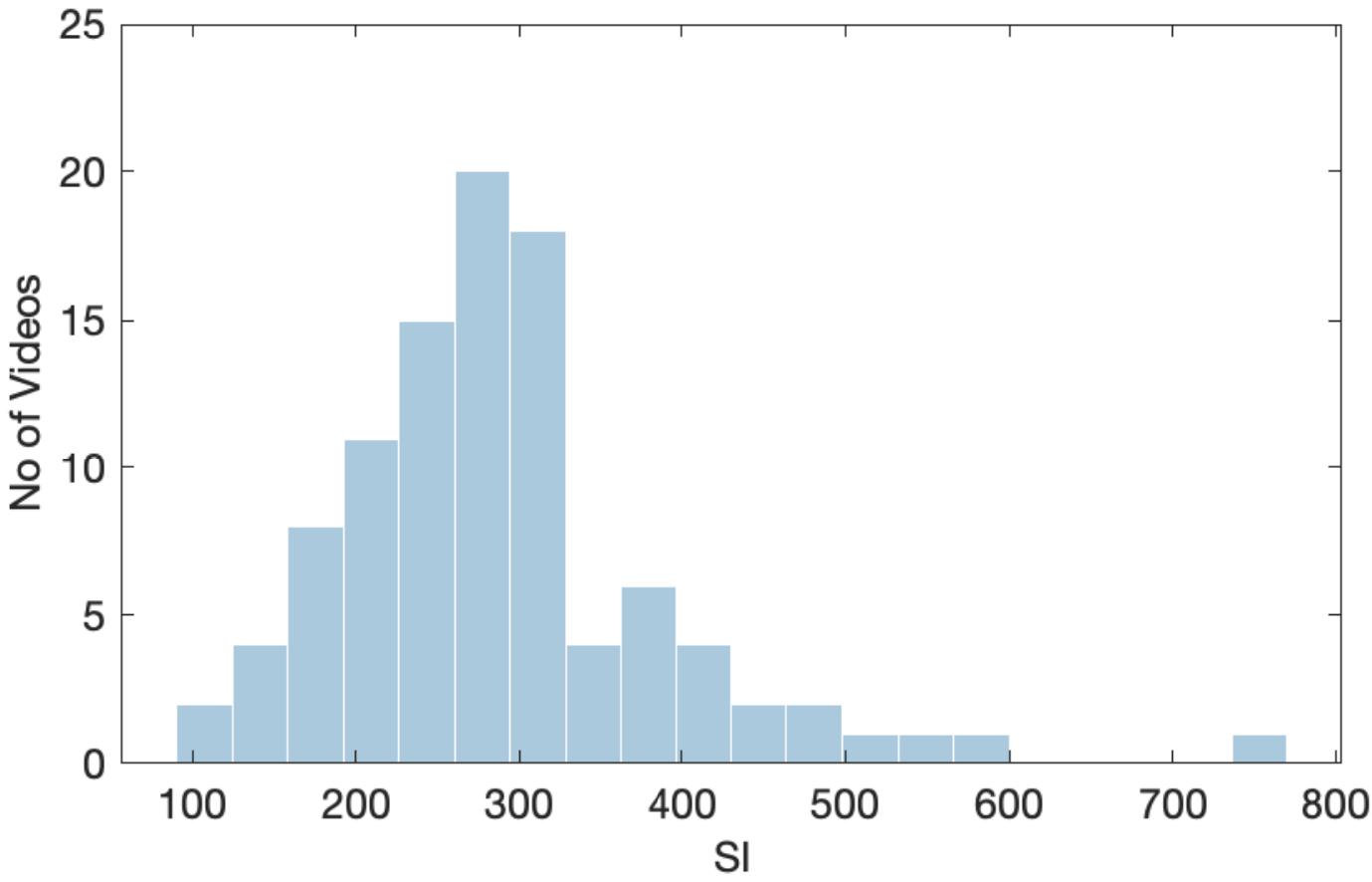}
			\vspace{-.5cm}
			\subcaption{SI histogram.}
		\end{minipage}
		\begin{minipage}[b]{.49\linewidth}
			\centering
			\includegraphics[width=\linewidth]{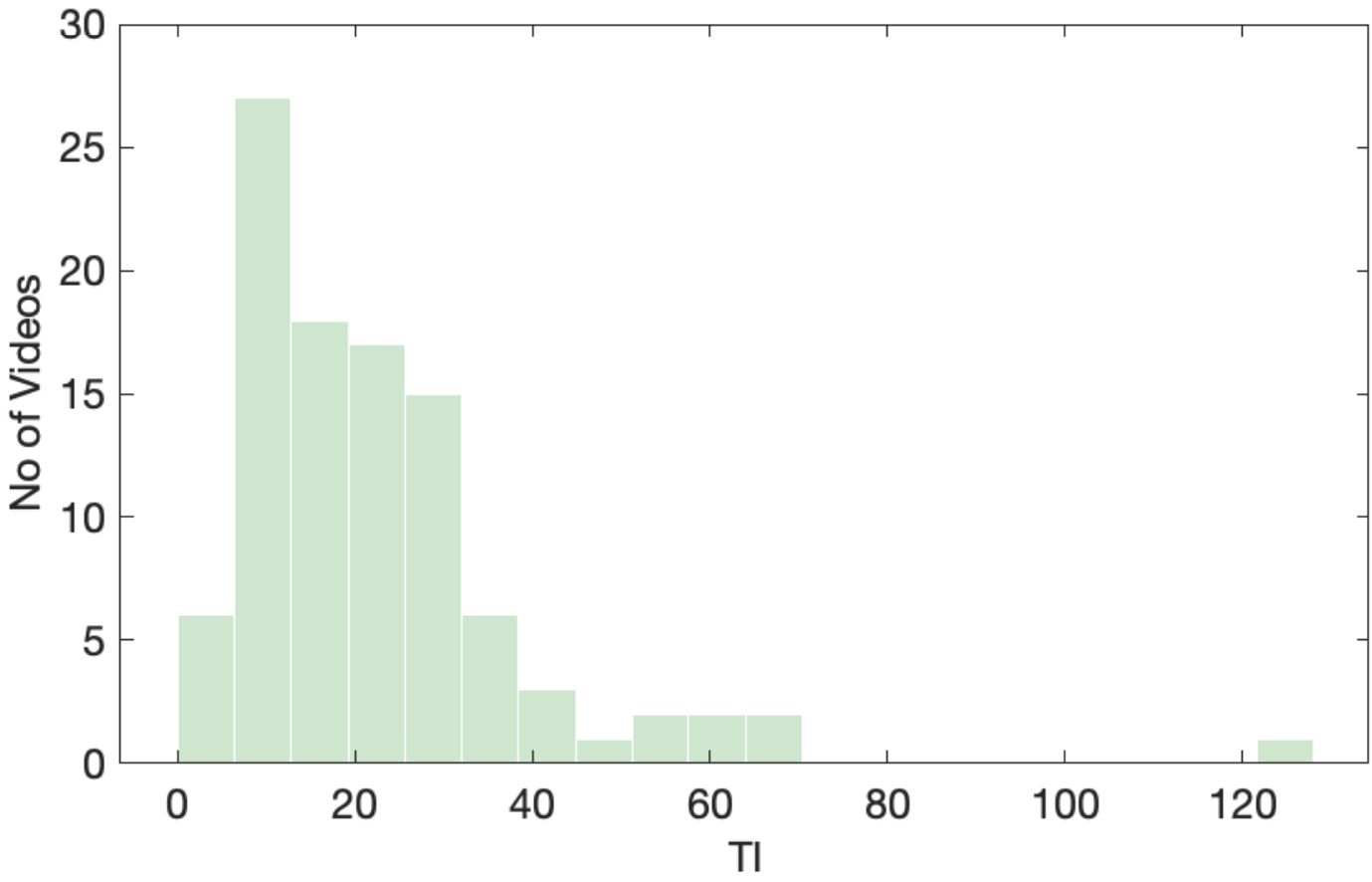}
			\vspace{-.5cm}
			\subcaption{TI histogram.}
		\end{minipage}
		\begin{minipage}[b]{.49\linewidth}
			\centering
			\includegraphics[width=\linewidth]{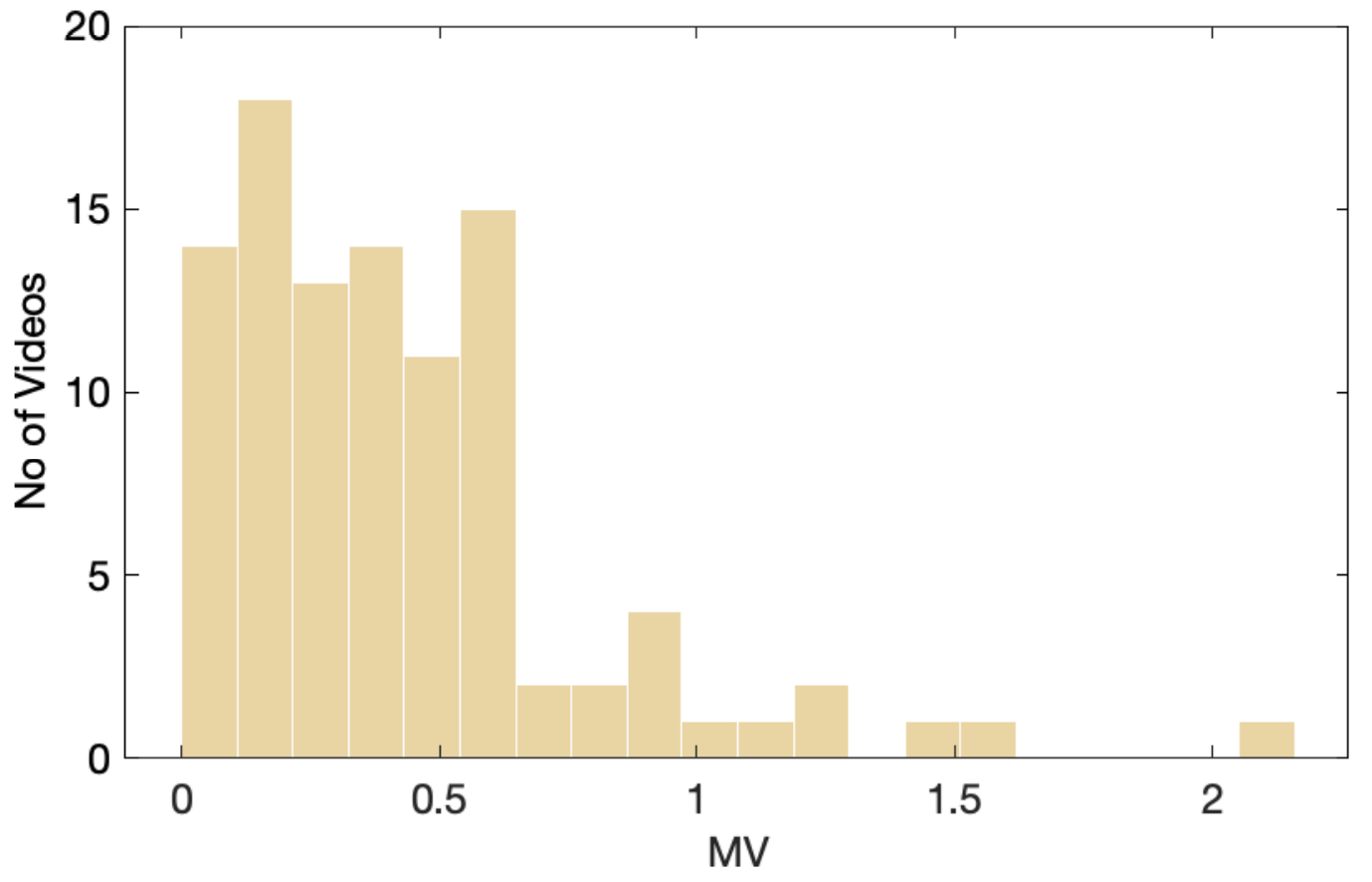}
			\subcaption{MV histogram.}
		\end{minipage}
		\begin{minipage}[b]{.49\linewidth}
			\centering
			\includegraphics[width=\linewidth]{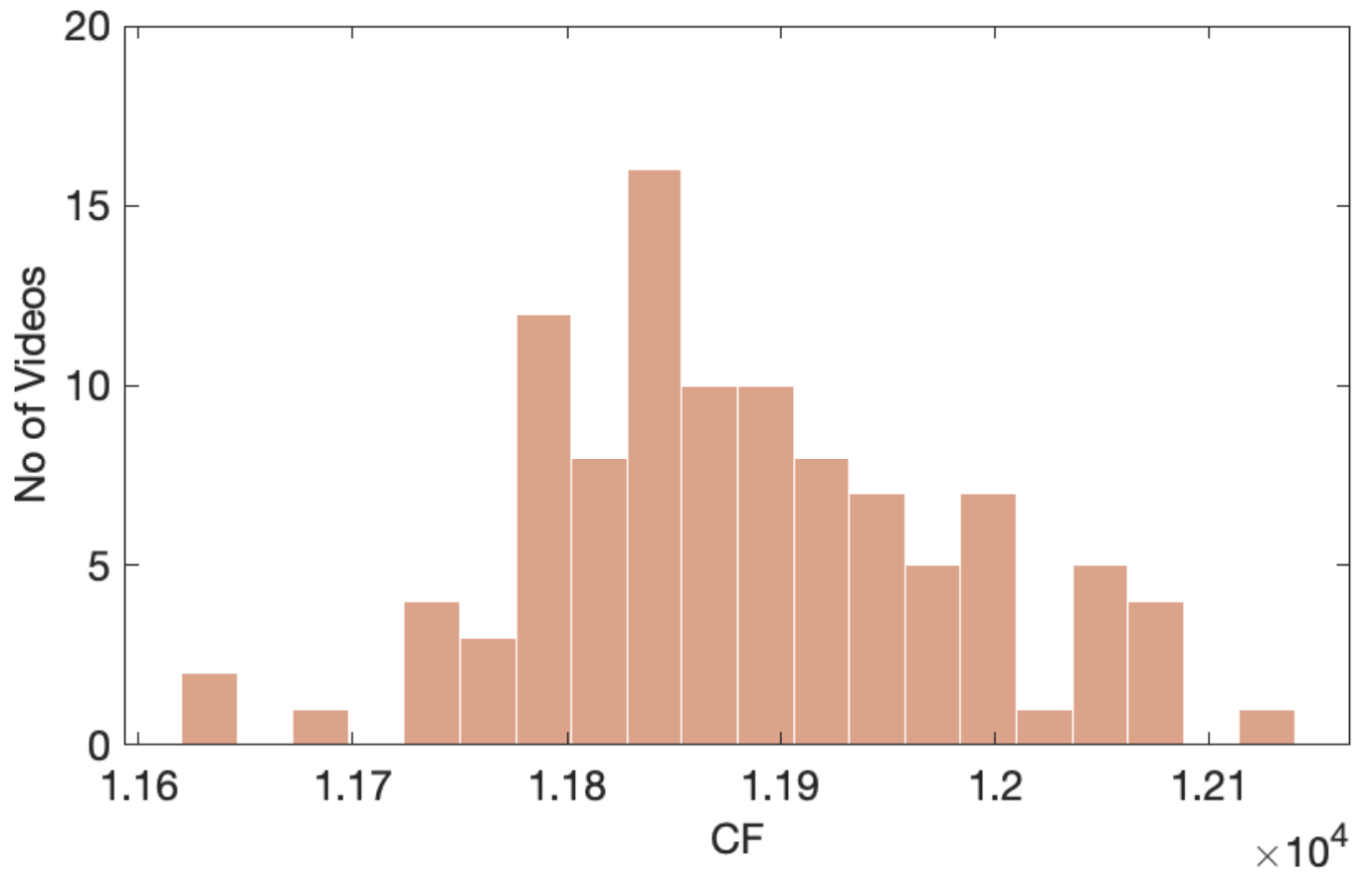}
			\subcaption{CF histogram}
		\end{minipage}
		\caption{Distributions of SI, TI, MV, and CF descriptors of the considered dataset.}
		\label{fig:SITIMV}
	\end{figure}

	We illustrate in Fig.~\ref{fig:SITIMV} the four basic descriptors of our dataset: Spatial Information (SI), Temporal Information (TI), average Motion Vectors (MV) magnitude and Colourfulness (CF)~\cite{winkler2012analysis}. These four distributions highlight the variety of its video content. SI is an indicator of edge energy; TI is an indicator of temporal variance; MV is another expression of how fast the motion in successive frames might be; and CF is an indication of the colour distribution. All features show a wide coverage of the spatio-temporal domain with most of the sequences in the range of 150-250 for SI and 10-50 for TI. 
	The histograms also reveal some of the outlying video sequences, such as TunnelFlag (sample frame in row 10 and column 1 of Fig.~\ref{fig:Sample100}) and Wood (row 10 and column 9) that contain very dense edges. Video sequence Jockey (row 3, column 6) also differs from others because of its high motion.
	
	\subsection{The Reference Pareto-optimal Front}
	\label{ssec: CHtheory}
	Each RQ curve at a resolution $S \in \mathcal{S}$ can be defined as the set of vectors; bitrate $\mathbf{R}=(R_1, R_2, \ldots, R_{|\mathbf{P}|})^\top$, video quality $\mathbf{Q}=(Q_1, Q_2, \ldots, Q_{|\mathbf{P}|})^\top$, and quantization points $\mathbf{P}=(P_1, P_2, \ldots, P_{|\mathbf{P}|})^\top$. As $P$ represents the independent parameter QP that is given as input parameter to the video codec, $\mathbf{R},\mathbf{Q}$ depend on $\mathbf{P}$. 
	
	Each RQ curve expresses the tradeoff between quality $Q$ and bitrate $R$ at a resolution $S$ over the independent parameter $P$. The sets of the independent variables $\mathcal{S}, \mathcal{P}$ form a decision variable space that is mapped to an objective function space that contains the resulting $R_i,Q_i$ points. Our aim is to determine the optimal $\langle P^*_i,S^*_i \rangle$ that result to the highest quality $Q^*_i$ at the lowest possible $R^*_i$. These tuples of optimal points $\{\langle R^*_i,Q^*_i,P^*_i,S^*_i \rangle \}$ form the PF. Every point on the PF is dominant over every other point in the objective function space. To put it simply, the PF, $\mathcal{C}$, in our work is defined as a set of tuples, i.e.
	\begin{equation}
	\label{eq: convhull}
	\mathcal{C}:=\{\langle R^*_i,Q^*_i,P^*_i,S^*_i \rangle \}^{|\mathcal{C}|}_i \; 
	\end{equation}
	where $R^*_i>R^*_{i-1}$ with $R_i \in \mathds{R}^{+,*}$ and $R_{\min}<R_i<R_{\max}$, $Q^*_i>Q^*_{i-1}$ with $Q_i \in \mathds{R}^{+,*}$, $P*_i \leq P^*_{i-1}$ with $P_i \in \mathds{N}^{*}$, $S*_i \geq S^*_{i-1}$, and $|\mathcal{C}|$ expressing the cardinality and varying per content. Depending on the shape of the RQ curves across resolutions and their intersection points, the PF shape and cardinality may differ per sequence.
	
\begin{figure}[!t]
		\begin{minipage}[b]{.49\linewidth}
			\centering
			\includegraphics[width=\linewidth]{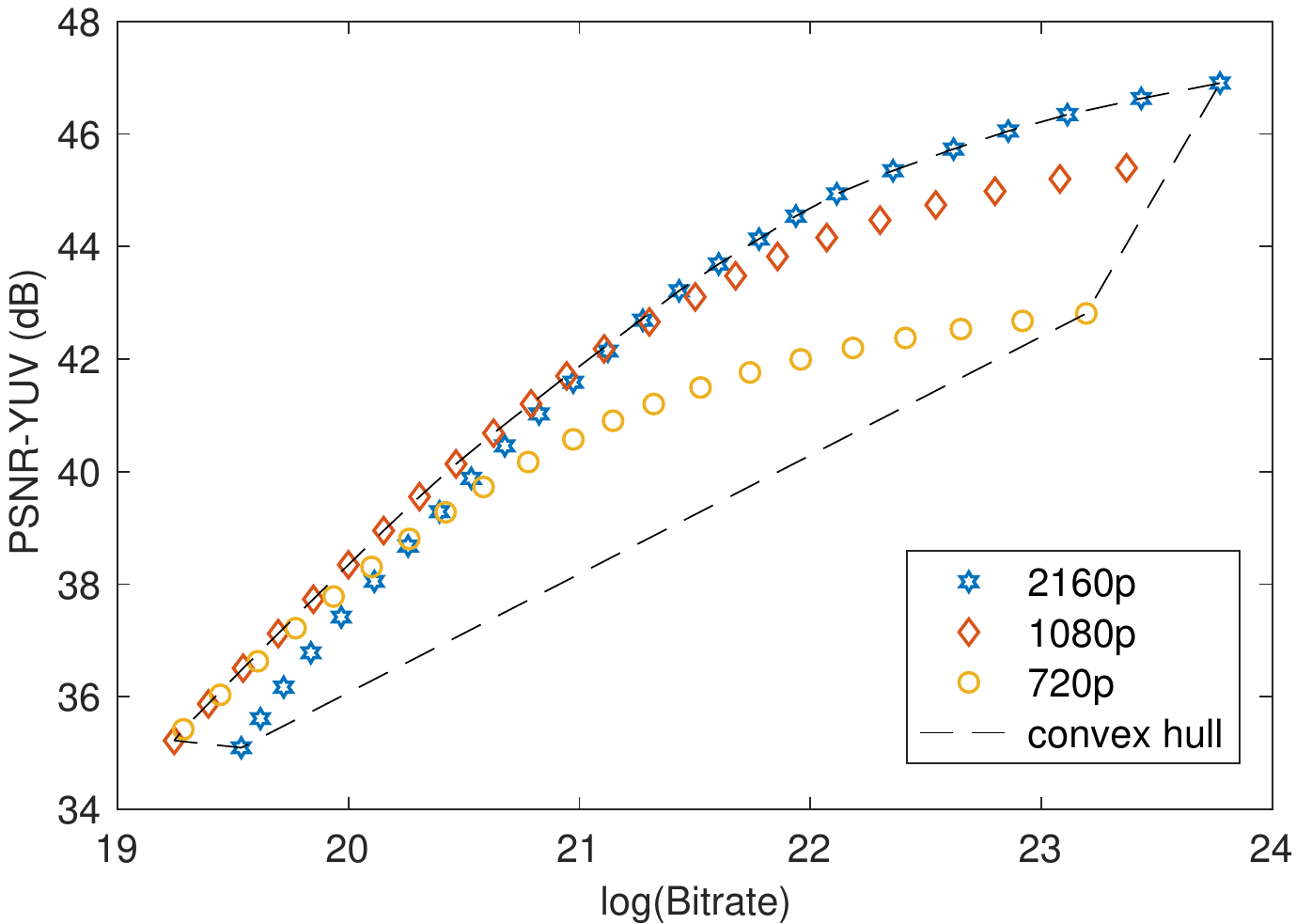}
			\subcaption{RQ points and the feasible objective space.}
		\end{minipage}
		\hfill
		\begin{minipage}[b]{.49\linewidth}
			\centering
			\includegraphics[width=\linewidth]{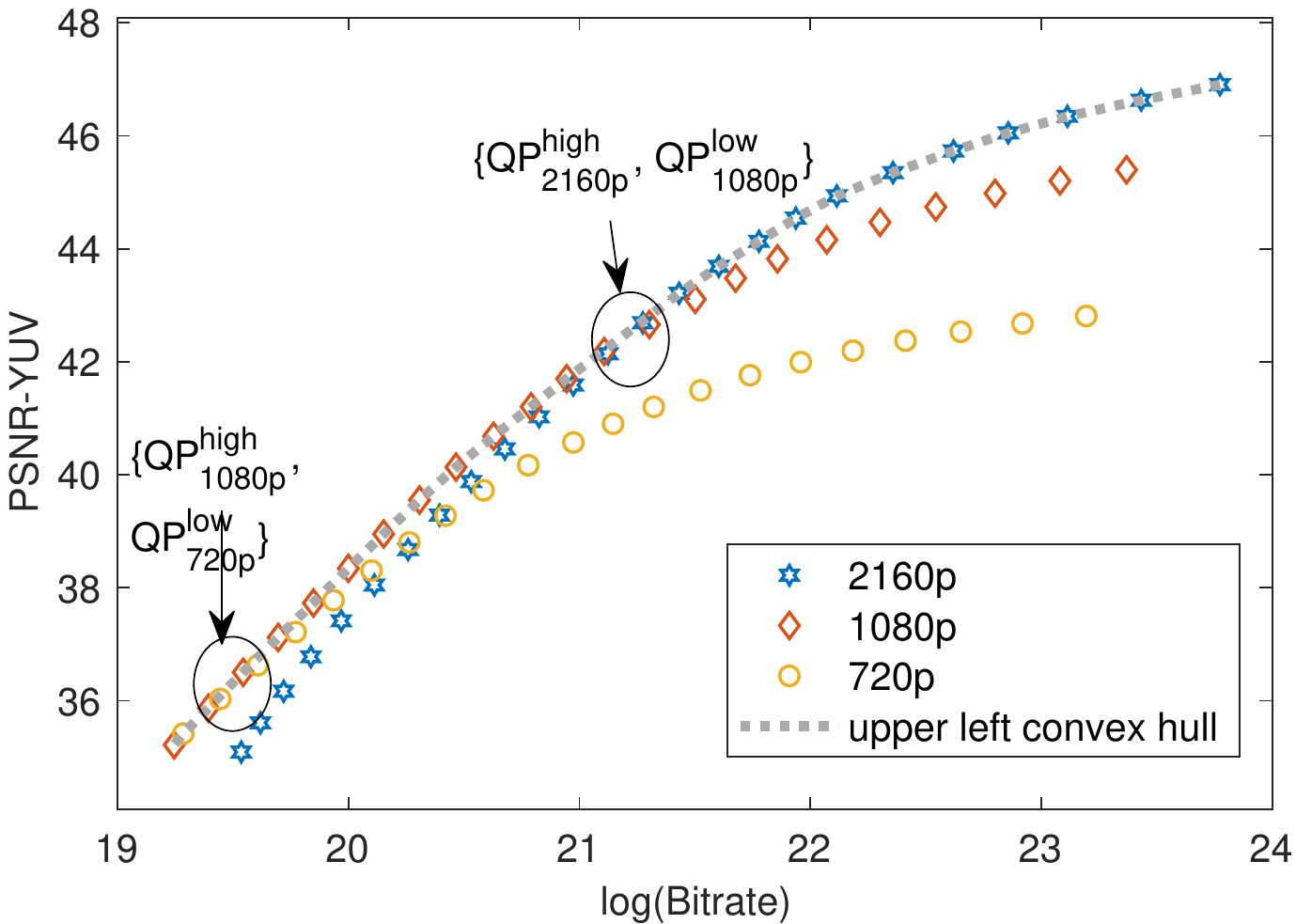}
			\subcaption{Intersections of RQ curves and cross-over QPs.}
		\end{minipage}
		\caption{Example of intersecting RQ curves at (2160p, 1080p and 720p), feasible objective space and cross-over points for the sequence Marathon.}
		\label{fig:exampleCH_crossQP}
	\end{figure}

	In Fig.~\ref{fig:exampleCH_crossQP} (a), an example of the Rate-PSNR points across three resolutions, i.e. \{2160p, 1080p, 720p\}, and the respective feasible objective space is depicted. Next, in Fig.~\ref{fig:exampleCH_crossQP} (b), the PF is illustrated with the grey dashed line.

	Furthermore, it is important to define the intersection points between the RQ curves of the same video across resolutions. The intersection points signal the switching of resolutions and  are defined by pairs of QPs, called cross-over QPs, that are mathematically represented as follows
	\[
	\left(QP^{level_j}_{S_j},QP^{level_{j-1}}_{S_{j-1}}\right), 
	\]
	with
	\[
	S_j \neq S_{j-1} \; , 
	\]
	\[    level_j \neq level_{j-1} \; ,
	\]
	where $S_j, j \in \{1,2,\ldots,|\mathcal{S}|\}$ are resolutions of the intersecting curves of the same video sequence and $level \in \{high, low\}$ defines the range of QPs. The total number of cross-over QPs depends of course on the number of resolutions and equals to $2\times(|\mathcal{S}|-1)$. This was used in order to make more distinct the intersection QPs of the same resolution. So, $level$ is an indication of whether the intersection is happening at high or low range of QP values. The resolution and level cannot be the same for both QPs in a pair. For example, the pair ($QP^{high}_{2160p},QP^{low}_{1080p}$) indicates that the 2160p curve is intersecting with the 1080p curve at a considerably high value for 2160p and at a low for 1080p. Figure~\ref{fig:exampleCH_crossQP} (b) illustrates an example of cross-over QPs and how the notation is used for the three intersecting curves. 
	
	\subsection{Constructing the Reference Pareto surface}
	\label{ssec: ConstrTheorCH}
	We first construct the ground truth PF and determine the intersection points of the RQ curves between different spatial resolutions. These intersection points mark the limits of the range for which encoding at the given resolution yields the best quality. When encoding at a lower resolution, all metrics are computed on the rescaled version (see Fig.~\ref{fig: method}): all sequences are first downscaled, then encoded, decoded and finally upscaled prior to metric computation.
	
	We spatially downscaled all sequences in our dataset (see Fig.~\ref{fig: method}) using a Lanczos-3 filter~\cite{Duchon}, as implemented by FFmpeg~\cite{ffmpeg}, at four different resolutions, $\mathcal{S}=\{2160p, 1080p, 720p, 540p\}$. Then, we encoded all versions of the sequences with the HEVC reference software, HM16.20~\cite{HEVCpaper}, using the Random Access profile, a 64-frame intra period, a length of group of pictures equal to 16 frames, and a fixed QP range for all resolutions: $\mathcal{P}=\{15, \ldots, 45\}$. The range of QPs selected was sufficiently wide to ensure that the RQ curves across resolutions intersect. As can be seen in Fig.~\ref{fig: method}, after decoding the sequences, we upscale them to the native resolution using the same filter. All quality metrics are computed at the display resolution (2160p), as recommended also in~\cite{Sole2018Netflix}.

	\begin{figure}[!t]
		
		\begin{minipage}[b]{.49\linewidth}
			\centering
			\includegraphics[width=\linewidth]{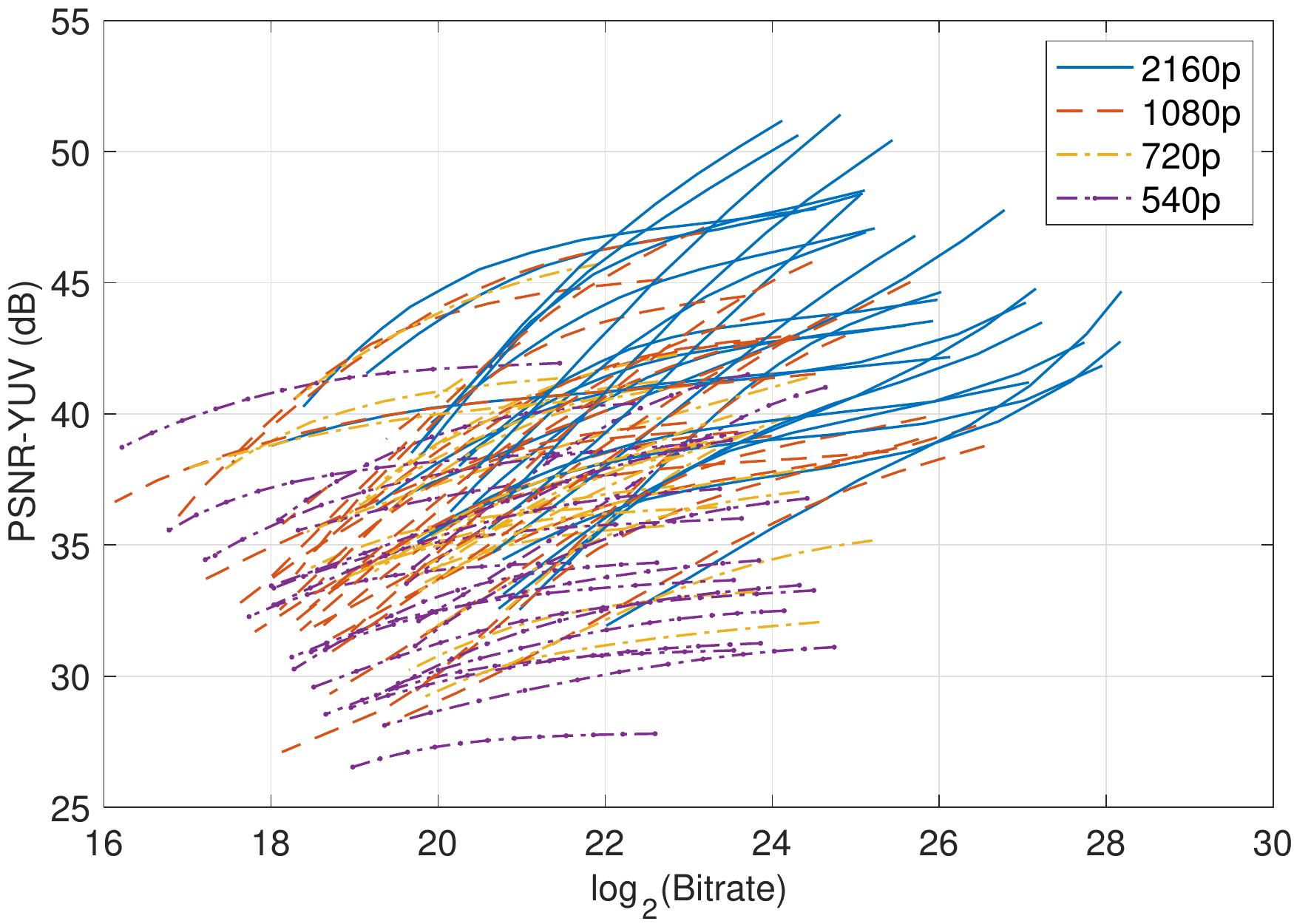}
			\footnotesize{(a) $\log$(R)-PSNR curves.}
		\end{minipage}
		\hfill
		\begin{minipage}[b]{.49\linewidth}
			\centering
			\includegraphics[width=\linewidth]{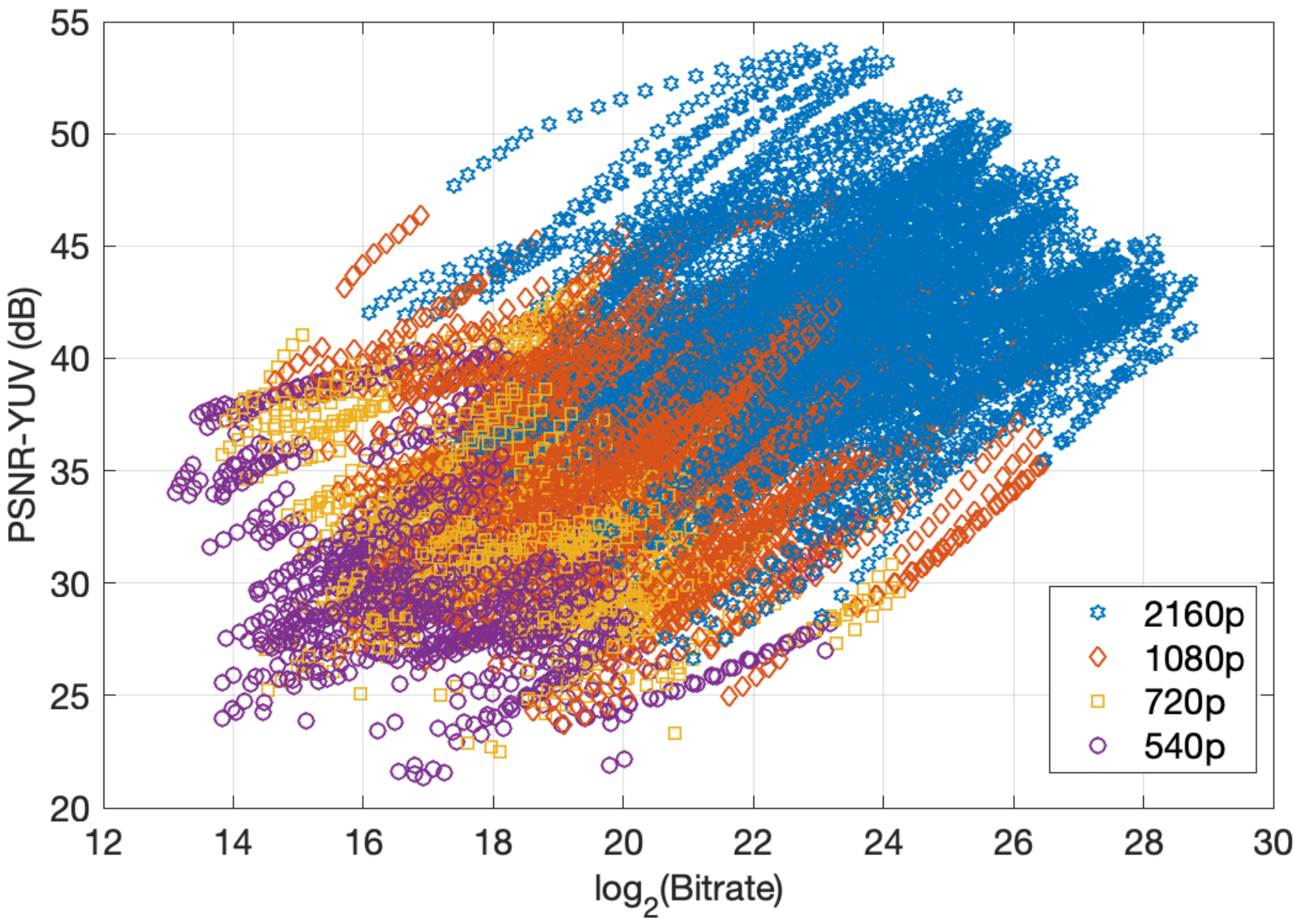}
			\footnotesize{(b) $\log$(R)-PSNR PFs.}
		\end{minipage}
		
		\caption{Examples of a subset of the considered dataset RQ curves at four different spatial resolutions.}
		\label{fig:CrossRQ}
	\end{figure}
	
	In Fig.~\ref{fig:CrossRQ}~(a), we illustrate a subset of  $\log_2$(Rate)-PSNR curves from the considered dataset across four spatial resolutions for the same range of quantization levels ($\log_2$ is henceforth simplified as $\log$). 
	From these figures, we observe that the wide range of content features is reflected by the high diversity in the RQ curves. For example, the sequence ToddlerFountain (row 5, column 7, Fig.~\ref{fig:Sample100}) that exhibits dynamic texture (fountains) has a smoother slope, compared to a more static sequence such as HoneyBee (row 7, column 8, Fig.~\ref{fig:Sample100}) that exhibits a steeper slope. Furthermore, there is a shift of the RQ curves toward lower quality and bitrate associated with downscaled spatial resolutions. The lower resolution sequences saturate at lower quality values. However, sequences at lower resolutions demonstrate higher quality values at lower bitrates. Also, the intersection points  differ significantly according to sequence characteristics (see Section~\ref{ssec: predCrossQP}, Fig.~\ref{fig: feat_vs_QP4K}).  
	In Fig.~\ref{fig:CrossRQ}~(b), we illustrate the resulting Pareto surfaces for our dataset across the four spatial resolutions. As expected, the composition of these curves varies for the different sequences. The PFs are composed by a different number of points across resolutions. This figure emphasizes the requirement for content-aware bitrate ladder construction.
	
	After computing quality metrics on all versions of upscaled decoded sequences, we construct the PF for each sequence. This is referred to as the reference PF and will be considered as the ground truth. We assume that two curves across resolutions can only intersect once, and in a monotonically descending manner: 2160p with 1080p, 1080p with 720p, etc. Hence, if the resulting PF reveals more than one intersection between pairs of resolutions, we record the one with the highest bitrate.

	\subsection{Cross-over QPs and their relationships}
	\label{ssec: CrossQPsRelation}
	In this paper, we consider four spatial resolutions $\mathcal{S}=\{2160p, 1080p, 720p, 540p\}$, so we have three intersections that correspond to three pairs of cross-over QPs, namely $(QP^{high}_{2160p}, QP^{low}_{1080p})$, $(QP^{high}_{1080p}, QP^{low}_{720p})$, and $(QP^{high}_{720p}, QP^{low}_{540p})$. A typical example of intersecting RQ curves is drawn in Fig.~\ref{fig:exampleCH_crossQP}~(b). We can see that the RQ curves across resolutions reside in very close proximity, appearing to overlap across a wide range of bitrates. Such occurrences are common and very often the quality values differ only marginally across resolutions for certain bitrate ranges. 
	
\begin{figure}[!t]
		\begin{minipage}[b]{.49\linewidth}
			\centering
			\includegraphics[width=\linewidth, trim={0 0 0 0},clip]{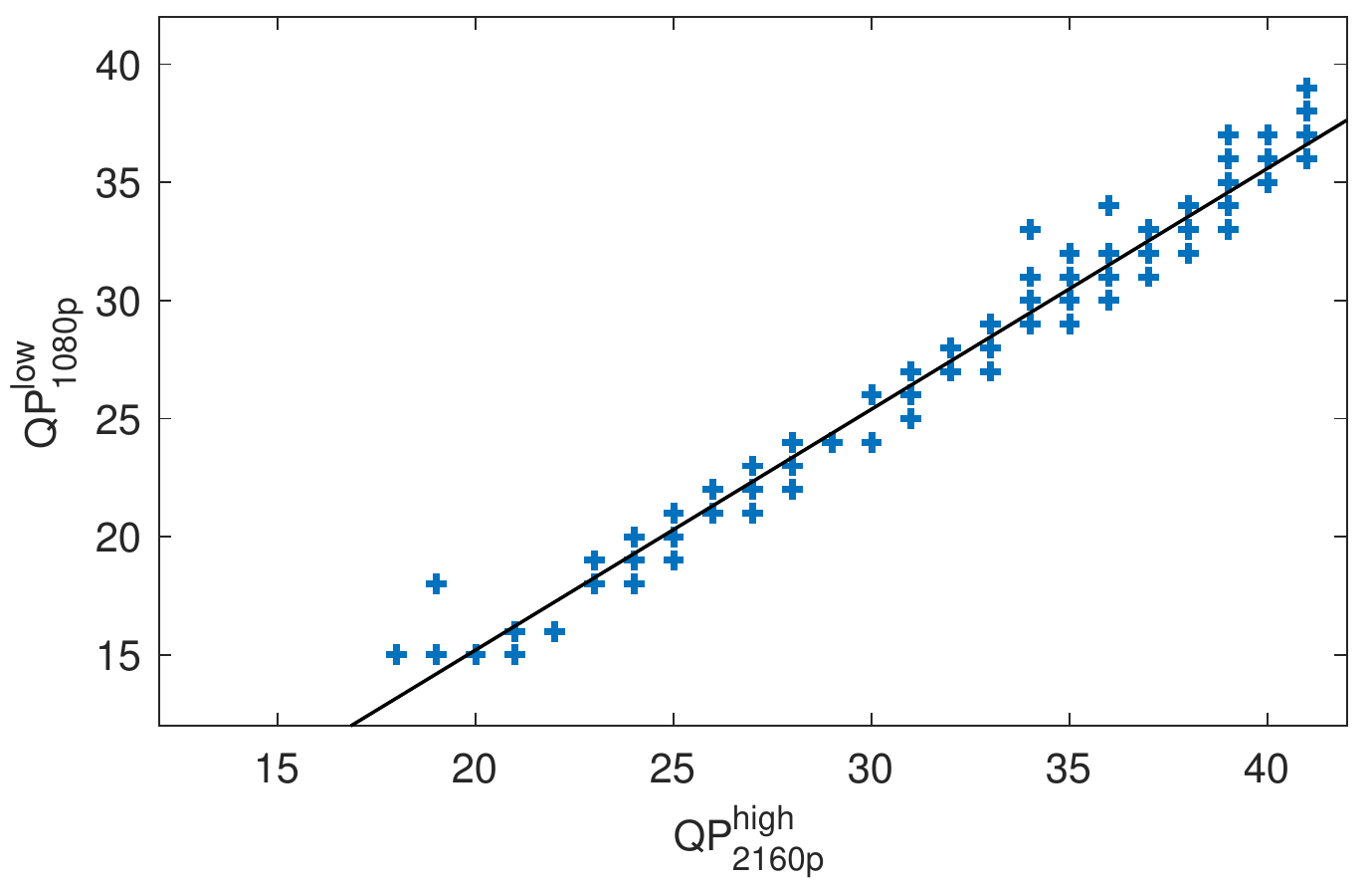}
			\subcaption{$QP^{low}_{1080p}$ vs $QP^{high}_{2160p}$.}
		\end{minipage}
		\begin{minipage}[b]{.49\linewidth}
			\centering
			\includegraphics[width=\linewidth, trim={0 0 0 0},clip]{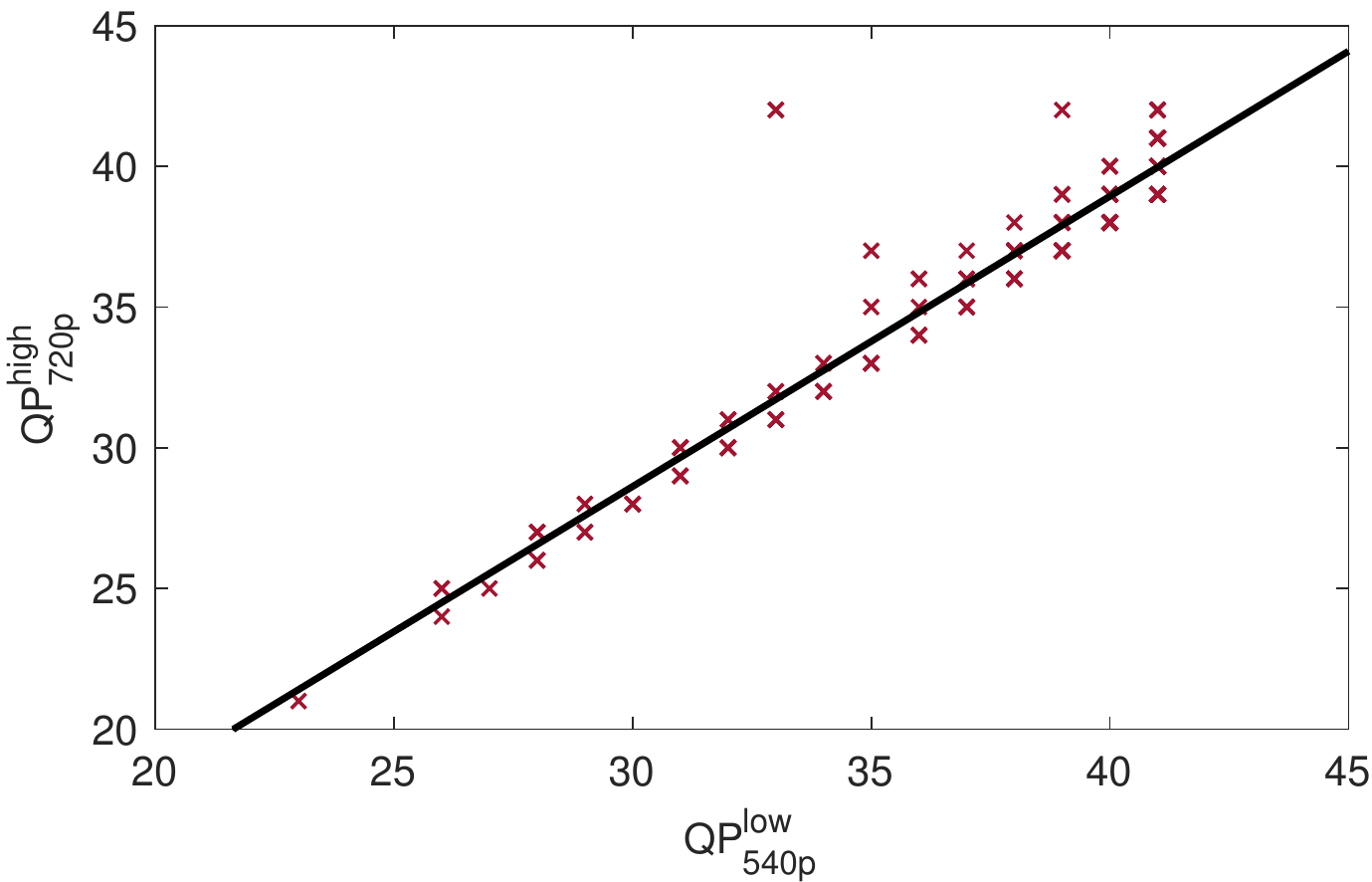}
			\subcaption{$QP^{high}_{720p}$ vs $QP^{low}_{540p}$.}
		\end{minipage}
		\caption{Examples of scatter plots of two cross-over QP pairs.}
		\label{fig:CrossQPs}
	\end{figure}

	Moreover, many of the cross-over QPs are highly correlated across different resolutions. 
	Table~\ref{tab: crossQPs_corr} reports the Pearson Linear Correlation Coefficient (LCC) and Spearman Rank Correlation Coefficient (SROCC) for the cross-over QPs. Almost all QPs are highly correlated. These observations are useful, indicating that previously predicted cross-over QPs could be used as features.
	The linear relationship between pairs of cross-over QPs can be verified from the example scatter plots in Fig.~\ref{fig:CrossQPs}, where two examples of pairs of cross-over QPs are given. It can be seen that the cross-over points show a close-to-linear shift across resolutions. 
	An example of this linear relationship between cross-over QPs is given below:
	\begin{equation}
	\widetilde{QP}^{low}_{1080p}=1.02 QP^{high}_{2160p}-5.17 \; ,
	\end{equation}
	\begin{equation}
	\widetilde{QP}^{high}_{720p}=1.03 QP^{low}_{540p}-2.30 \; ,
	\end{equation}
	where the estimated QP values are rounded to the nearest integer. In this case, LCC is 0.9908 and SROCC 0.9901 for the  $(QP^{high}_{2160p}, QP^{low}_{1080p} )$ pair and 0.9563 and 0.9160 for the other pair, respectively. 
	
	\begin{table*}[!ht]
		\caption{Cross-correlation values (LCC, SROCC) between cross-over QPs.} 
		\centering
		\begin{tabular}{l|cccccc}
			\toprule
			\diagbox{}{PSNR}&$QP^{high}_{2160p}$& $QP^{low}_{1080p}$ & $QP^{high}_{1080p}$ & $QP^{low}_{720p}$ & $QP^{high}_{720p}$ & $QP^{low}_{540p}$\\
			\midrule
			$QP^{high}_{2160p}$& - & (0.9908, 0.9901)  & (0.8974, 0.8638) & (0.9108, 0.8825) & (0.7908, 0.7414) & (0.8378, 0.8073)\\
			$QP^{low}_{1080p}$& - &-  & (0.8842, 0.8521)  & (0.9040, 0.8812)  & (0.7700, 0.7318) & (0.8149, 0.7971)\\
			$QP^{high}_{1080p}$ & - & - &-  & (0.9895, 0.9825)  & (0.9123, 0.8865) & (0.9140, 0.8690)\\
			$QP^{low}_{720p}$& - & - & - &-  & (0.9000, 0.8728) & (0.9182, 0.8760)\\
			$QP^{high}_{720p}$& - & - & -  & - &- & (0.9563, 0.9160)\\
			\bottomrule
		\end{tabular}
		\label{tab: crossQPs_corr}
	\end{table*}
	
	\subsection{The Reference Bitrate Ladder}
	\label{ssec: BuildLadder}
	After constructing the Pareto-optimal front, the next step is to build the bitrate ladder. We define the bitrate ladder as an ordered set $\mathcal{R_L}=\{R_{L,1},R_{L,2},\ldots,R_{L,|\mathcal{L}|}\}$, where $|\mathcal{L}|$ is the cardinality of $\mathcal{R_L}$ and $R_{L,1}<R_{L,2}<\ldots<R_{L,N}$. The bitrate ladder is fully defined as a set of tuples $\mathcal{L}$ that comprise bitrate values $\mathcal{R_L}$,  the associated set of quality values $\mathcal{Q_L}$,  a set of QP values $\mathcal{P_L}$, and  a set of resolutions $\mathcal{S_L}$, i.e.
	\begin{equation}
	\label{eq: Ladder}
	\mathcal{L}:= \{\langle R_{L,i},Q_{L,i},P_{L,i},S_{L,i} \rangle \}^{|\mathcal{L}|}_i \; .
	\end{equation}
	
	In order to construct the bitrate ladder, we follow three steps. First, we define the range of bitrates that will be used for streaming, and trim our PF to lie between the lower $R_{\min}$ and upper $R_{\max}$ bitrate values. Next, we perform subsampling of the trimmed front across bitrate and quality. 
	
	Constraints across the quality dimension depend on the metric employed. From Fig.~\ref{fig:CrossRQ} (b), we observe that, for some sequences, the PF saturates after reaching a certain bitrate value. Allocating bits beyond this value would not improve video quality. A shorter bitrate ladder, that takes into account the saturation for these sequences can therefore be used. In general, as mentioned in Section~\ref{sec:Intro}, the length of the ladder will depend on the video content and its compression performance across the different resolutions.
	
	Next, we follow common practise by selecting points on the PF such that each ladder point $R_{\textrm{L,}i}$ is approximately twice the bitrate of the previous point, i.e. 
	\begin{equation}
	R_{\textrm{L,}i} \simeq 2R_{\textrm{L,}i-1}  ,
	\end{equation}
	where $R_{\textrm{L,}i} \in \left(R_{\min}, R_{\max}\right)$ and $i\in\mathds{N}$. This expression translated into the $\log_2$ domain can be written as
	\begin{equation}
	\label{eq: trimCH}
	\log (R_{\textrm{L,}i}) \simeq 1+\log (R_{\textrm{L,}i-1})
	\end{equation}
	We use approximation in the above equation because, in practice the curves are not continuous, but instead finite sets of discrete points as a consequence of using integer QP values.
	
	We next subsample the PF considering restrictions across the quality dimension. 
	Put formally, we find the rate points on the ladder $R_{\textrm{L,}i}$ for which:
	\begin{equation}
	\label{eq: trimCH_Q1}
	Q_{\textrm{L,}i}(R_{\textrm{L,}i})  \leq Q_{\max} \; , \; 
	\end{equation}
	\begin{equation}
	\label{eq: trimCH_Q2}
	\diff{Q_{\textrm{L,}i}}{R_{\textrm{L}}} > \epsilon \; ,
	\end{equation}
	where $Q_{\max}$ is the maximum value that can be assumed by normalised metrics and $\epsilon\in \mathds{R}, \epsilon \rightarrow 0$. 
	As a consequence of the above constraint, the length of the ladder might vary. The use of different ladder lengths, dependent on  compression complexity was also suggested in~\cite{Reznik2018}.
	The basic steps to construct the reference bitrate ladder explained above are briefly outlined in Algorithm~\ref{alg: algorithm1}.
	
	\begin{algorithm}[!h]
		\SetAlgoLined
		\KwIn{video sequence, set of resolutions $\mathcal{S}$, set of quantization points $\mathcal{P}$}
		\KwOut{Reference bitrate ladder $\mathcal{L}$ per video sequence}
		\vspace{7px}
		
\% \textit{\textbf{Step1:} Extract RQ Points}\\
		\For{each $s\in\mathcal{S}$}{
			Downscale sequence at $s$ using Lanczos-3 filter\;
			\For{each $p\in\mathcal{P}$}{
				Encode sequence at QP$=p$ with RA profile, intraPeriod$=64$, GoPlength$=16$\;
				Compute Bitrate, $R_p$\;
				Decode sequence\;
				Upscale decoded sequence to native resolution\;
				Compute quality metrics $Q_p$\;
			}
			RQ curve at $s, \{ \mathbf{\log(R)},\mathbf{Q},\mathbf{P},\mathbf{S} \}$
		}
		\Return RQ curves across all resolutions, $\{ \mathbf{\log(R)},\mathbf{Q},\mathbf{P},\mathbf{S} \}$
		\vspace{7px}
		
\% \textit{\textbf{Step2:} Compute the Reference PF}\\
		Find the RQ points that compose the PF\;
		Find the intersection points of the RQ curves\;
		\Return Reference PF: $\mathcal{C}_{ref} \gets \{\langle \log(R_i),Q_i,P_i,S_i \rangle \}^{|\mathcal{C}|}_i $.
		\vspace{7px}
		
\% \textit{\textbf{Step3:} Compute the Reference Bitrate Ladder}\\
		Trim the logarithmic bitrate range: $\log(R)_{\min} \leq \log(R)_{\textrm{L,}i} \leq \log(R)_{\max}$ \;
		Prune the trimmed $\mathcal{C}$ across bitrate dimension using Eq.(\ref{eq: trimCH}), $\longrightarrow \mathcal{C'}$ \;
		Prune $\mathcal{C'}$ across the quality dimension according to Eqs.(\ref{eq: trimCH_Q1})-(\ref{eq: trimCH_Q2}) \;
		
		\vspace{7px}
		
		\Return Reference Bitrate Ladder , as in~Eq.(\ref{eq: Ladder}): $\mathcal{L} \gets \{\langle \log(R)_{L,i},Q_{L,i},P_{L,i},S_{L,i} \rangle \}^{|\mathcal{L}|}_i$.
		
		\caption{Reference Bitrate Ladder Construction}
		\label{alg: algorithm1}
	\end{algorithm}
	
\begin{figure}[!h]
		\begin{minipage}{.49\linewidth}
			\centering
			\includegraphics[width=\linewidth]{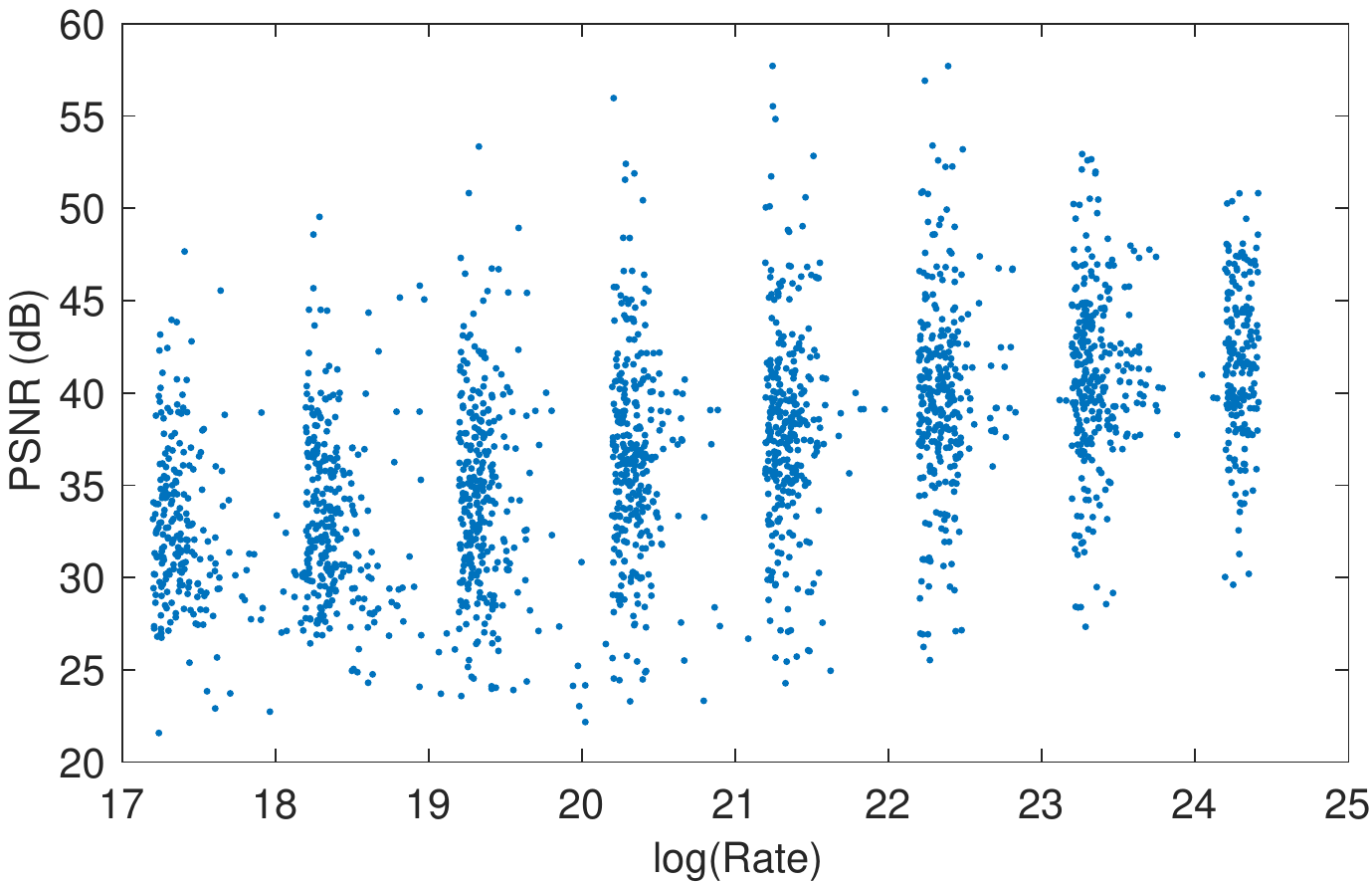}
			\footnotesize{(a) Bitrate ladder points.}
		\end{minipage}
		\begin{minipage}{.49\linewidth}
			\centering
			\includegraphics[width=\linewidth]{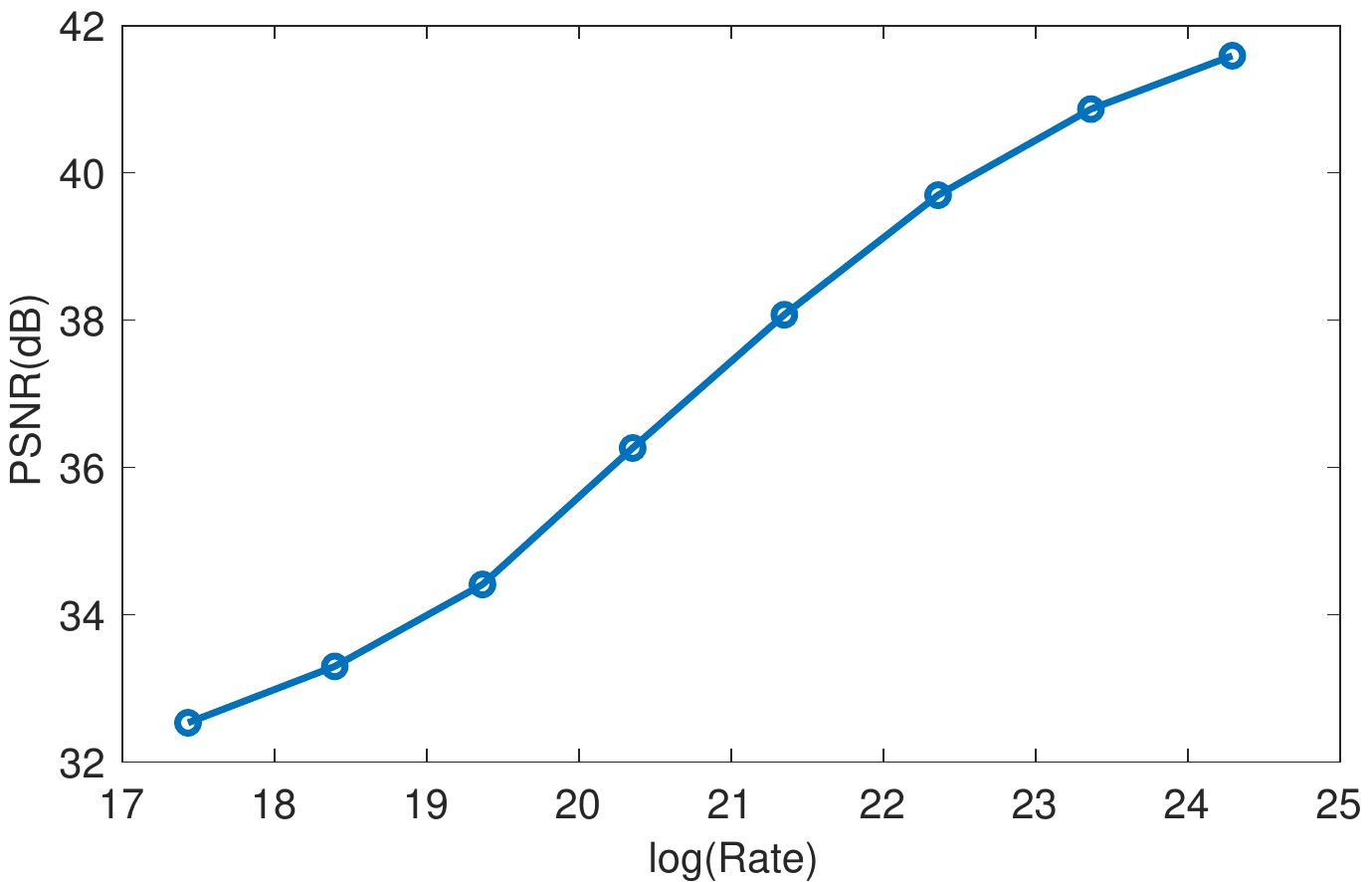}
			\footnotesize{(b) Average Bitrate ladder.}
		\end{minipage}
		\caption{The reference bitrate ladder for the considered dataset.}
		\label{fig:BirateLadder}
	\end{figure}

	In Fig.~\ref{fig:BirateLadder}, the reference bitrate ladders (a) per sequence and (b) on average for the considered dataset, are illustrated. For these figures, we considered the \{150kbps,25Mbps\} bitrate range. As can be seen from both plots, the steps on the ladder are clearly visible and are shifted to a greater or lesser extent according to the sequence. It is noticeable in Fig.~\ref{fig:BirateLadder}~(a) that the last step of the bitrate ladder appears to have a smaller number of points. This indicates the variable length of the ladder as a result of the PF in the considered bitrate range. 
	
	Table~\ref{tab:CHpatterns} reports the statistics on the different combinations detected in the considered dataset. As expected, the combinations of higher resolutions are dominant. It can be seen, that only for 13.69\% of the test sequences, the native resolution is not included in the constructed reference bitrate ladder.
	The content-gnostic construction of a bitrate ladder offers the advantage of combining a variable length bitrate ladder and lower than native resolutions where appropriate, as opposed to the traditional method that includes all resolutions and all target ladder rungs. This results in reduced encoding cost, while not degrading the end-user experience.

	\begin{table}[!ht]
		\caption{Resolution patterns that compose the bitrate ladders.} 
		\centering
		\begin{tabular}{cccc|r}
			\toprule
			\textbf{2160}&\textbf{1080} & \textbf{720}& \textbf{544}& \textbf{PSNR}\\
			\midrule
			\checkmark& \checkmark& \checkmark& \checkmark& 44.13\%\\
			\checkmark& \checkmark& \checkmark& -&  17.32\%\\
			\checkmark& \checkmark& -& -&  21.79\%\\
			\checkmark& -& -& -& 3.07\%\\
			-& \checkmark& \checkmark& \checkmark&  11.73\%\\
			-& \checkmark& \checkmark& -& 0.56\%\\
			-& \checkmark& -& -&  1.4\%\\
			\bottomrule
		\end{tabular}
		\label{tab:CHpatterns}
	\end{table}

	\begin{algorithm}[!h]
		\SetAlgoLined
		\KwIn{test video sequence $v$ at native resolution}
		\KwOut{Predicted bitrate ladder $\mathcal{\Tilde{L}}$ per video sequence}
		
\% \textit{\textbf{Step1:} Extract Features}\\
		\For{each frame $i\in \{1, 2, \ldots, NoFrames\}$}{
			Compute GLCM contrast, homogeneity, correlation, energy, entropy on frame $i$\;
			\uIf{$i>1$}{ 
				Compute TC mean, standard deviation, skewness, kurtosis and entropy between frames $(i-1,i)$\;
			}
			\ElseIf{$i=1$}{
				Compute RsMSE\;}
		}
		Compute the mean GLCM descriptors, the mean and standard deviation of TC statistics over all frames\;

\% \textit{\textbf{Step2:} Predict Cross-Over QPs}\\
		\For{each $QP^{level}_s$}{
			Select a subset of features using RFE\;
			Predict $\widehat{QP}^{level}_s$\;
			Update the set of features with the $\widehat{QP}^{level}_s$\;
		}
		
\% \textit{\textbf{Step3:} Estimate the QP-$\log$(R) Eq. Parameters }\\
		\For{each $v$}{
			\For{each $s\in\mathcal{S}$}{
				\uIf{$s\neq|\mathcal{S}|$}{
					\If{$s>1$}{
						Downscale sequence $v_d$ at resolution $s$ using Lanczos-3 filter\; 
					}
					Encode sequence $v_d$ at $ \widehat{QP}^{high}_{s}$\;
				}
				\Else{
					Encode sequence $v$ at $ \widehat{QP}^{low}_{s}$\;
					
				}
				Compute Bitrate, $R_p$\;
				Decode sequence $v'$\;
				\If{$s>1$}{
					Upscale sequence $v'$ at native resolution using Lanczos-3 filter\; 
				}
				Compute quality metrics $Q_p$ between $(v',v)$\;
			}
			Estimate Eq.(\ref{eq: QP(R)}) parameters for video $v'$ for all $s$\;
		}
		
\% \textit{\textbf{Step4:} Compute the Bitrate Ladder}\\
		Average the $R_p$ points of cross-over QPs to define the resolution switching bitrate.\\
		Repeat Lines 20-21 from Algorithm~\ref{alg: algorithm1} for each $v'$\;
		
\% \textit{\textbf{Step5:} Validate Monotonicity and Concavity}\\
		Order non-monotonic points and remove concave points.
		
		\Return Predicted Bitrate Ladder: $\mathcal{\widehat{L}} \gets \{\langle \log(R)_{L,i},\widehat{Q}_{L,i},\widehat{P}_{L,i},\widehat{S}_{L,i} \rangle \}^{|\mathcal{L}|}_i$.
		
		\caption{Prediction of the Bitrate Ladder}
		\label{alg: algorithm2}
	\end{algorithm}
	
	\section{Content-driven Prediction of the Bitrate Ladder}
	\label{sec: ConvHullPred}
	
	This section outlines the processes linked to the prediction of the PF, including feature extraction,  feature selection,  prediction of the cross-over points using machine learning models, estimation of the PF parameters and the assessment and evaluation of the results. A description of the proposed methodology for content-aware ladder prediction is given in Algorithm~\ref{alg: algorithm2}. Furthermore, this section includes a discussion of the results with respect to the optimality of the predicted bitrate ladders and the computational cost in terms of required encodings.
	
	\subsection{Spatio-Temporal Features}
	In this subsection, we discuss the spatio-temporal feature extraction, which is the first step for the prediction of the content-gnostic bitrate ladder, as described in Algorithm~\ref{alg: algorithm2}.
	
	Observing the RQ curves, their intersection points and their PFs, it is evident that there are strong dependencies on, and correlation with, content characteristics. For example, in the case of the Marathon sequence, whose PSNR-$\log$(R) curves are depicted in Fig.\ref{fig:exampleCH_crossQP}, we observe that the PF comprises many 2160p resolution points. This can be attributed to the density of small moving structures (runners) within the scene. On the other hand, for other more static sequences that include out of focus background (e.g. Barscene), the 1080p curve intersect with the 2160p occurs at a much lower QP value. The challenge is to find suitable spatio-temporal features that reflect such content characteristics.
	
	The literature is rich with various spatio-temporal features used to characterise the relationship between video content and compression performance\cite{MPEG7book,Haralick1973, Haralick1979, PappasTIP2013, KaramICIP15, DelpJSTSP2011, ZhangJSTSP2011, PehTIP2002}.
	The spatio-temporal features employed in this work have been carefully selected through extensive evaluation of a large variety of features, modifying some so that they better represent the basic characteristics of video texture that relate to encoding difficulty, i.e. spatial diversity, coarseness and motion, as shown in~\cite{KatsenouMMSP2017}. 
	
	We adopt those features that have been successfully used in our previous compression-related research~\cite{KatsenouPCS2016,KatsenouMMSP2017, KatsenouPCS2018, AfonsoSPIE2018}. Particularly, for the representation of spatial information, and specifically to express the variability of intensity contrast between neighboring pixels, we employ the Gray Level Co-occurrence Matrix (GLCM)~\cite{Haralick1973} and extracted its basic descriptors (contrast; correlation; homogeneity; energy; entropy) along with its mean and standard deviation across frames, as described in~\cite{KatsenouMMSP2017}. Another low-cost feature adopted is the Mean Squared Error of the spatial Rescaling (RsMSE) of the first frame, similarly to a feature suggested in~\cite{AfonsoSPIE2018}. This feature captures the distortions that result from spatial sub/upsampling. Furthermore, in order to combine both spatial and temporal characteristics, we employ Temporal Coherence (TC)~\cite{KatsenouPCS2016} with its interframe statistics: mean; standard deviation; skewness; kurtosis; and entropy, as well as the mean and standard deviation across all frames. Table~\ref{tab:feat} reports the full set of features and their statistics, biasing those with the lowest computational complexity and those that were  selected via feature selection methods as described in Section~\ref{sec: ConvHullPred}. Besides those referred to above, we have tested other features, including the normalised Laplacian pyramid~\cite{balle2016}, the normalised cross-correlation across successive frames~\cite{lewis1995fast,KatsenouPCS2016}, the average frame difference, the optical flow~\cite{farneback2003}, and more. 
	
\begin{figure}[!t]
		\begin{minipage}[b]{.49\linewidth}
			\centering
			\includegraphics[width=\linewidth]{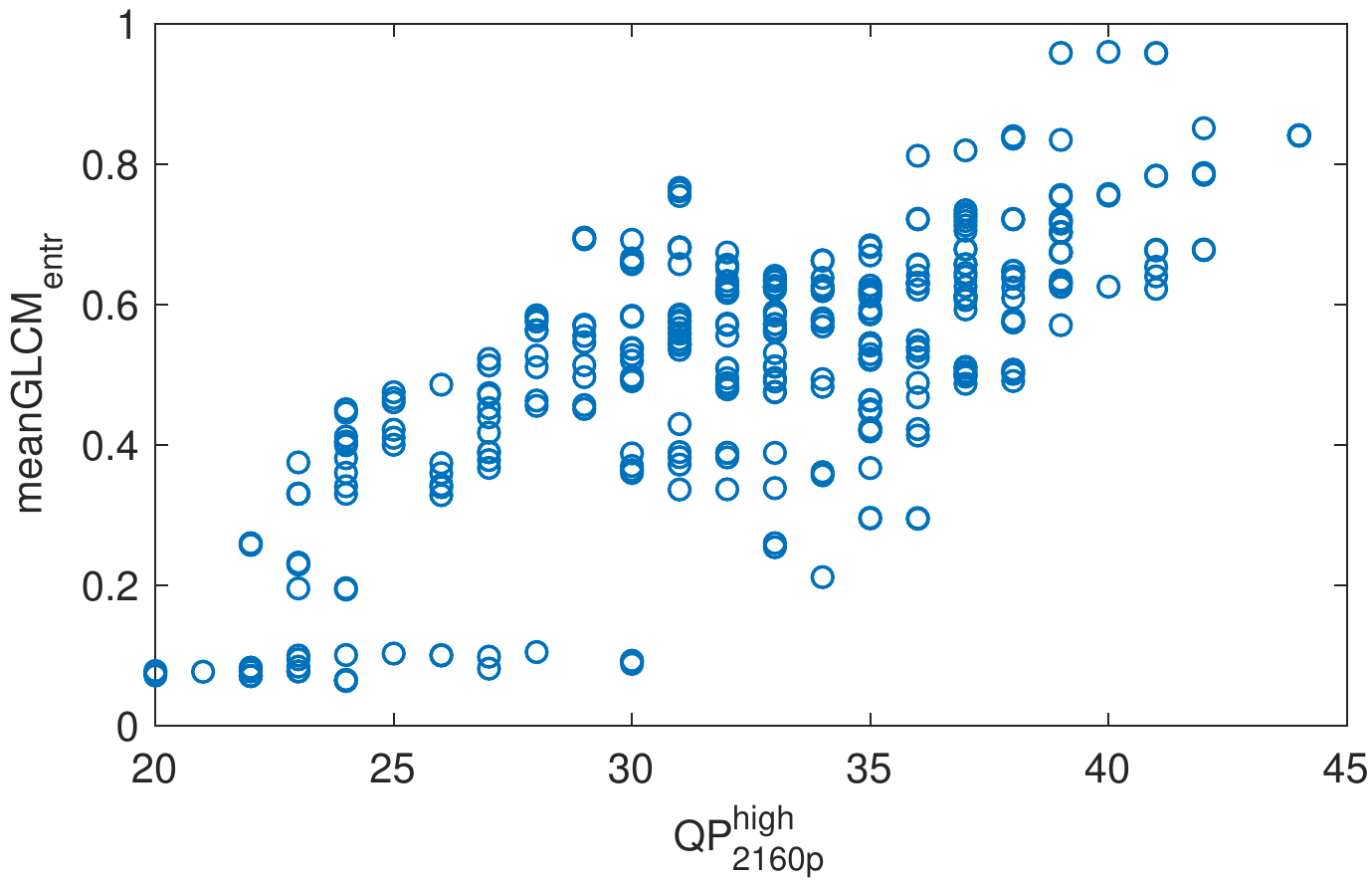}
			\footnotesize{(a) $QP_{2160p}$ vs $\textrm{meanGLCM}_{\textrm{ent}}$.}
		\end{minipage}
		\begin{minipage}[b]{.49\linewidth}
			\centering
			\includegraphics[width=\linewidth]{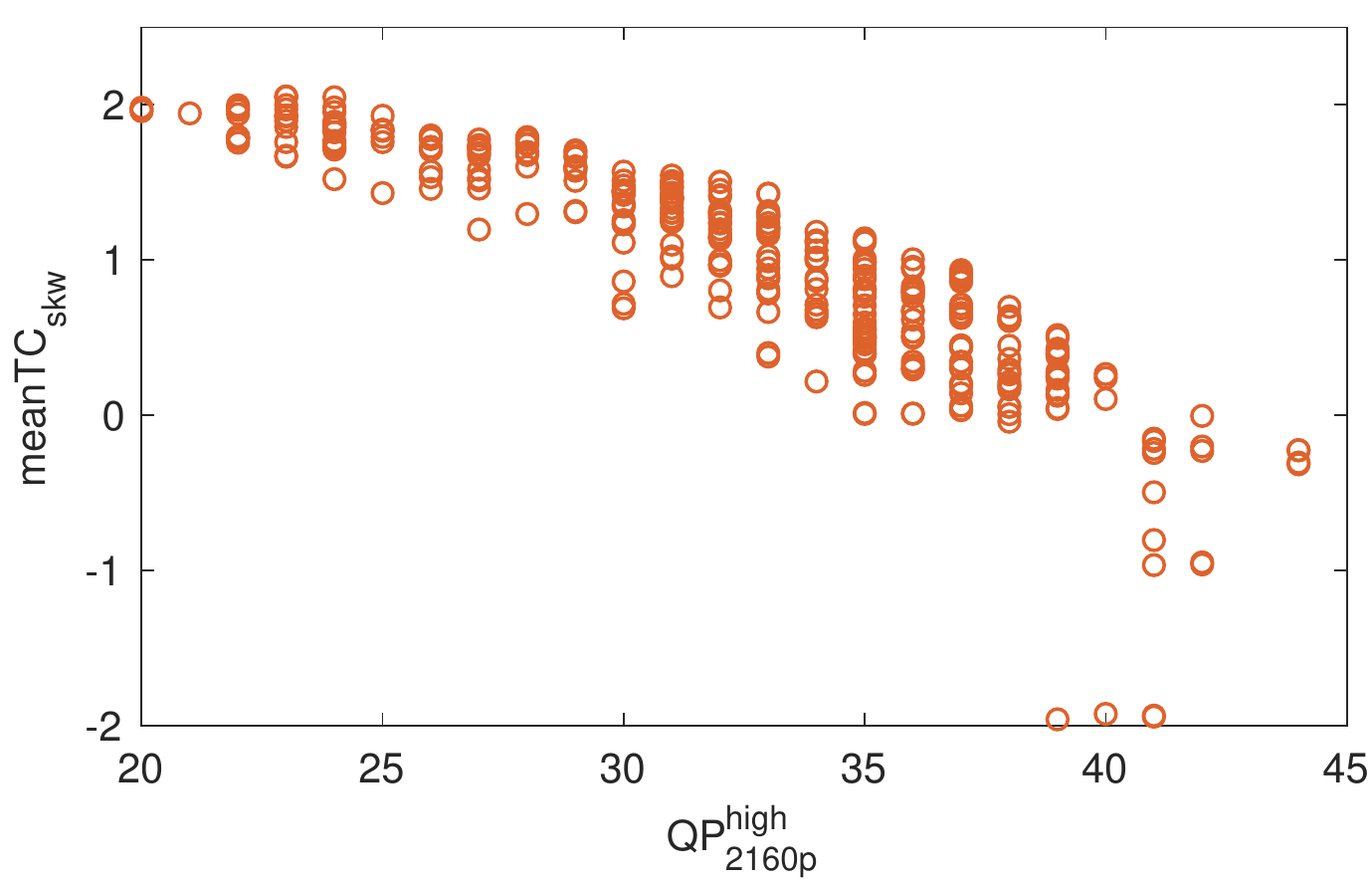}
			\footnotesize{(b) $QP_{2160p}$ vs $\textrm{meanTC}_{\textrm{skw}}$.}
		\end{minipage}
		\caption{Example of content dependency of the cross-over QPs.}
		\label{fig: feat_vs_QP4K}
	\end{figure}

	In Fig.~\ref{fig: feat_vs_QP4K}, we illustrate examples of the ground truth cross-over $QP^{high}_{2160p}$ against the temporal mean GLCM entropy, $\textrm{meanGLCM}_{\textrm{ent}}$, and the mean of the temporal coherence skewness, $\textrm{meanTC}_{\textrm{skw}}$, extracted from the video sequences at their native resolution. The higher the value of $\textrm{meanGLCM}_{\textrm{ent}}$, the higher the spatial variability of the sequence. This is, in most cases, related to a high cross-over value for the $QP^{low}_{2160p}$, which means that the PF comprises more points from the 2160p resolution. In the case of $\textrm{meanTC}_{\textrm{skw}}$, high values (positive skewness) indicate a temporally coherent sequence where switching to a lower resolution is likely to happen at a lower QP value. 
	
	It is also worth highlighting that, in the last row of Table~\ref{tab:feat}, the predicted cross-over QPs are listed as features. As explained in Section~\ref{ssec: CrossQPsRelation}, there is a strong relationship between cross-over QPs. There inclusion has resulted in higher prediction accuracy. More details on this follow in the next subsection.
	
	\begin{table}[!h]
		\begin{center}
			\caption{List of features and their notations.} \label{tab:feat}
			\footnotesize
			\begin{tabular}{l|l}
				\toprule
				\multicolumn{1}{p{3cm}|} {\textbf{Feature}} & \multicolumn{1}{p{4.2 cm}} {\textbf{Notations}}\\
				\cmidrule{1-2}
				\multicolumn{1}{p{2.7cm}|}  {Grey-Level Co-occurrence Matrix (GLCM)~\cite{Haralick1973}} & \multicolumn{1}{p{4.5 cm}}{\raggedright F1.$\textrm{meanGLCM}_{\textrm{con}}$, F6.$\textrm{stdGLCM}_{\textrm{con}}$, F2.$\textrm{meanGLCM}_{\textrm{cor}}$, F7.$\textrm{stdGLCM}_{\textrm{cor}}$, F3.$\textrm{meanGLCM}_{\textrm{hom}}$, F8.$\textrm{stdGLCM}_{\textrm{hom}}$, F4.$\textrm{meanGLCM}_{\textrm{enr}}$, F9.$\textrm{stdGLCM}_{\textrm{enr}}$, F5.$\textrm{meanGLCM}_{\textrm{ent}}$, F10.$\textrm{stdGLCM}_{\textrm{ent}}$} \\
				\cmidrule{1-2}
				\multicolumn{1}{p{2.7 cm}|} {Temporal Coherence (TC)~\cite{KatsenouPCS2016}}  &
				\multicolumn{1}{p{4.5 cm}}{\raggedright F11.$\textrm{meanTC}_{\textrm{mean}}$, F16.$\textrm{stdTC}_{\textrm{mean}}$, F12.$\textrm{meanTC}_{\textrm{std}}$, F17.$\textrm{stdTC}_{\textrm{std}}$, F13.$\textrm{meanTC}_{\textrm{skw}}$, F18.$\textrm{stdTC}_{\textrm{skw}}$, F14.$\textrm{meanTC}_{\textrm{kur}}$, F19.$\textrm{stdTC}_{\textrm{kur}}$, F15.$\textrm{meanTC}_{\textrm{entr}}$, F20.$\textrm{stdTC}_{\textrm{entr}}$ }\\
				\cmidrule{1-2}
				\multicolumn{1}{p{2.7 cm}|} {MSE from Rescaling using Lanczos filter (RsMSE)~\cite{AfonsoSPIE2018}} &
				\multicolumn{1}{p{4.5 cm}}{\raggedright F21.$\textrm{RsMSE}_\textrm{1080p}$ (from 2160p to 1080p), F22.$\textrm{RsMSE}_\textrm{720p}$ (from 2160p to 720p), F23.$\textrm{RsMSE}_\textrm{720p}$ (from 2160p to 720p)}\\
				\cmidrule{1-2}
				\multicolumn{1}{p{2.7 cm}|} {Predicted cross-over QPs} &
				\multicolumn{1}{p{4.5 cm}}{\raggedright F24.$\hat{QP}^{low}_{2160p}$, F25.$\hat{QP}^{low}_{1080p}$, F26.$\hat{QP}^{high}_{1080p}$, F27.$\hat{QP}^{low}_{720}$, F28.$\hat{QP}^{high}_{720p}$} \\
				\bottomrule
			\end{tabular}
		\end{center}
	\end{table}

	\subsection{Predicting the Cross-Over QP Values}
	\label{ssec: predCrossQP}
	The latest version of HEVC reference software, HM16.20, was employed in this study. The Random Access profile was used according to the Common Testing Conditions~\cite{HEVCpaper,CTC2017}, namely a 64-frame Intra Period and a Group of Pictures (GoP) length  of 16 frames. After encoding, decoding, and upscaling the spatial resolution to 2160p, we computed the quality metrics and bitrate at a GoP level. Computing the RQ curves at a GoP level enabled a larger coverage of the RQ space.
	
	Prior to prediction of each cross-over QP value, we applied feature selection, using recursive feature elimination~\cite{Kuhn2013}, on the set of spatio-temporal features. We followed a sequential prediction of the cross-over QPs starting from the highest resolution down to the lowest. Despite the fact that for the $QP^{high}_{4K}$ prediction we only relied on spatio-temporal features extracted from the uncompressed 2160p videos, for the rest of the cross-over QPs, we made use of their identified relations as explained earlier. We trained and tested several machine-learning regression methods, including Support Vector Machines with different kernels and Random Forests. We also evaluated deep-learning based regression with dense sequential layers (rectified unit activation and Adam optimiser). However, the Gaussian Processes (GP) classifier performed best for this work, as also shown in~\cite{KatsenouPCS2019}. To avoid overfitting, we deployed a ten-fold random cross-validation process. 
	
	The results in Table~\ref{tab:PredxQP_PSNR} report the outcome of the ten-fold cross-validation with the accuracy of prediction metrics averaged over the ten folds. The table also lists the selected features and accuracy of prediction for each predicted cross-over QP.
	Regarding the selected features subsets, we observe that there are similarities for all predicted cross-over QPs.
	Additionally, the previously predicted QPs were also selected as features, as explained earlier. 
	The two tables report high values of R$^2$, around 0.9. Moreover, the cross-correlation metrics LCC and SROCC between the predicted and the ground truth QPs are also high\footnote{We have to note that the predicted values were rounded to the nearest integer and clipped to the range of QP values, before computing the correlation metrics.}. Also, the Mean Absolute Error (MAE) and the Root Mean Squared Error (RMSE) are considerably low and comparable for all predicted cross-over QPs. It is important to point out that the effectiveness of the method cannot be fully assessed by these results;  the predicted cross-over QPs will be utilised to estimate the resolution switching bitrates and define models to estimate the bitrate ladder. Hence, the comparison of the predicted bitrate ladder to the reference will provide the full assessment of this framework.
	
	\begin{table*}[!htb]
		\caption{Selected features \& validation metrics of predicted cross-over QPs for PSNR-$\log$(R) curves.} 
		\centering
		\footnotesize
		\begin{tabular}{l|l|ccccc}
			\toprule
			\textbf{QP} & \textbf{Selected Features} & \textbf{LCC}& \textbf{SROCC}& \textbf{R$^2$}&\textbf{MAE} & \textbf{RMSE}\\
			\midrule
			$\widehat{QP}^{high}_{2160}$  & F2, F4, F5, F11, F12, F14 & .9350& .9164 & .91&1.41&  1.96\\
			\midrule
			$\widehat{QP}^{low}_{1080}$  & F2, F4, F5, F11, F12, F14, F24 & .9442& .9296& .90& 1.35& 1.96\\
			\midrule
			$\widehat{QP}^{high}_{1080}$  & F2, F4, F5, F11, F12, F14, F25  & .9536& .9076& .91& .95& 1.36\\
			\midrule
			$\widehat{QP}^{low}_{720}$  & F2, F4, F5, F11, F12, F14, F21, F25, F26 & .9531& .8751&.92 & .76& 1.15\\
			\midrule
			$\widehat{QP}^{high}_{720}$ &  F2, F4, F5, F12, F13, F14 & .9316& .8334&.88 & .97& 1.44\\
			\midrule
			$\widehat{QP}^{low}_{544}$  &  F2, F4, F5, F11-F15 & .9210& .8535&.89 & 1.17& 1.59\\
			\bottomrule
		\end{tabular}
		\label{tab:PredxQP_PSNR}
	\end{table*}

	\subsection{Modelling QP-log(Rate) to Estimate the Bitrate Ladder}
	\label{ssec: QPRateModel}
\begin{figure}[!t]
		\begin{minipage}{.48\linewidth}
			\centering
			\includegraphics[width=\linewidth]{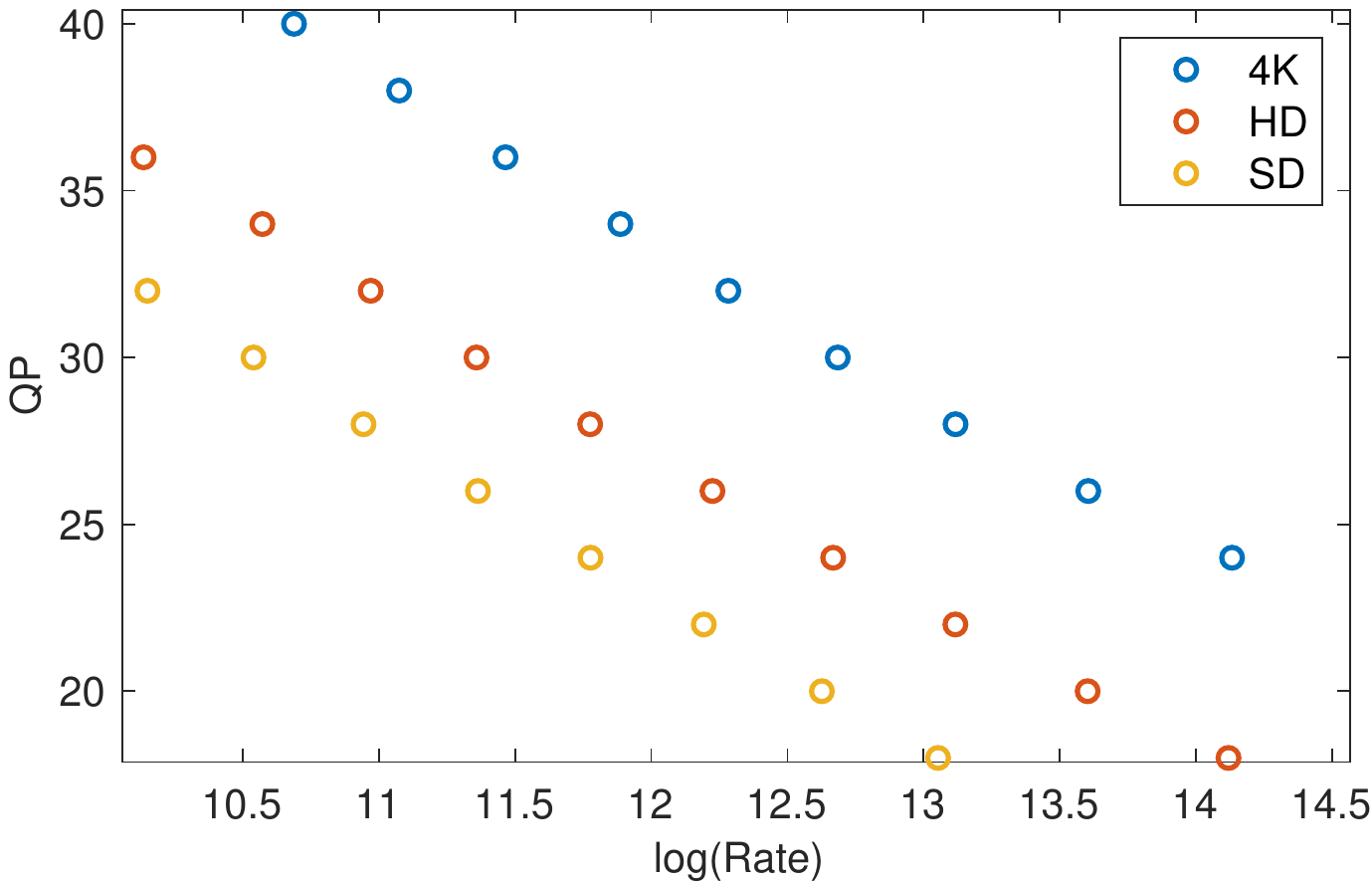}
			\footnotesize{(a) QP vs $\log$(R) across resolutions for ToddlerFontain.}
		\end{minipage}
		\hfill
		\begin{minipage}{.48\linewidth}
			\centering
			\includegraphics[width=\linewidth]{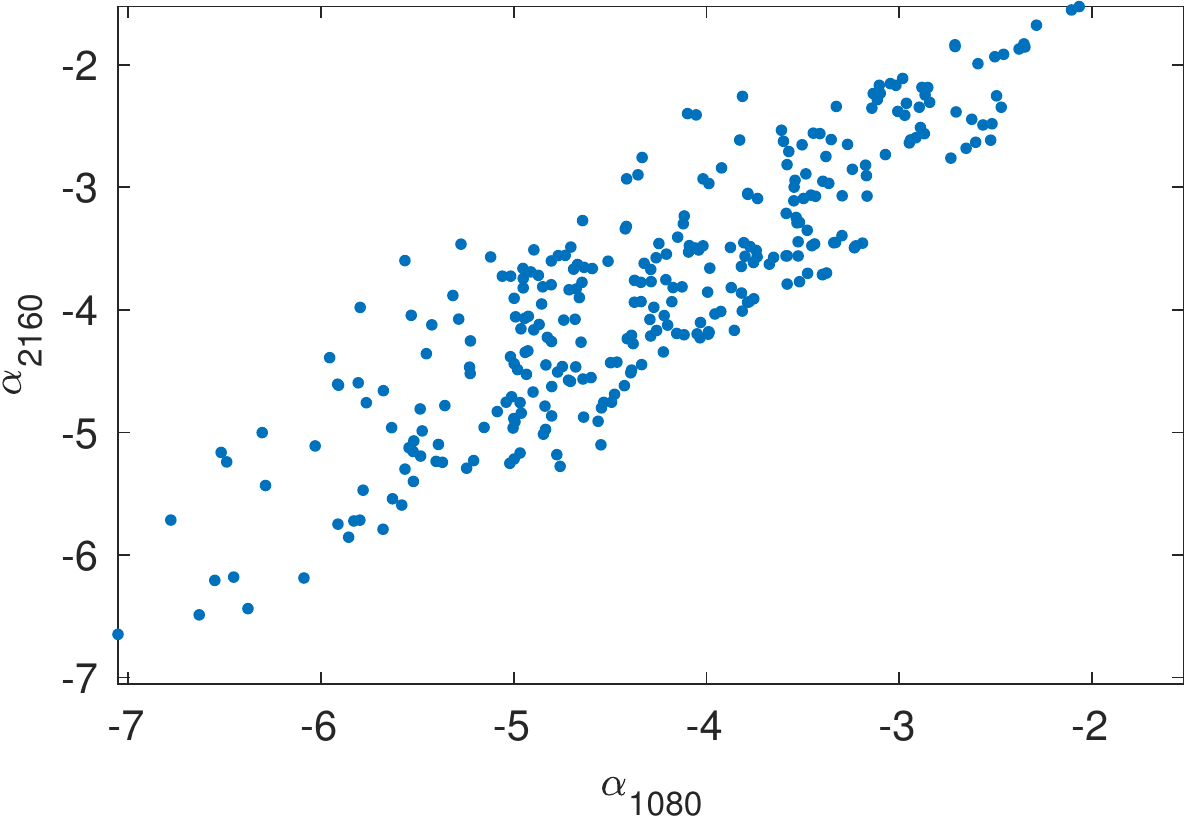}
			\footnotesize{(b) $\alpha_{2160p}$ vs $\alpha_{1080p}$ (LCC: 0.8748, SROCC: 0.8560).}
		\end{minipage}
		\hfill
		\begin{minipage}{.48\linewidth}
			\centering
			\includegraphics[width=\linewidth]{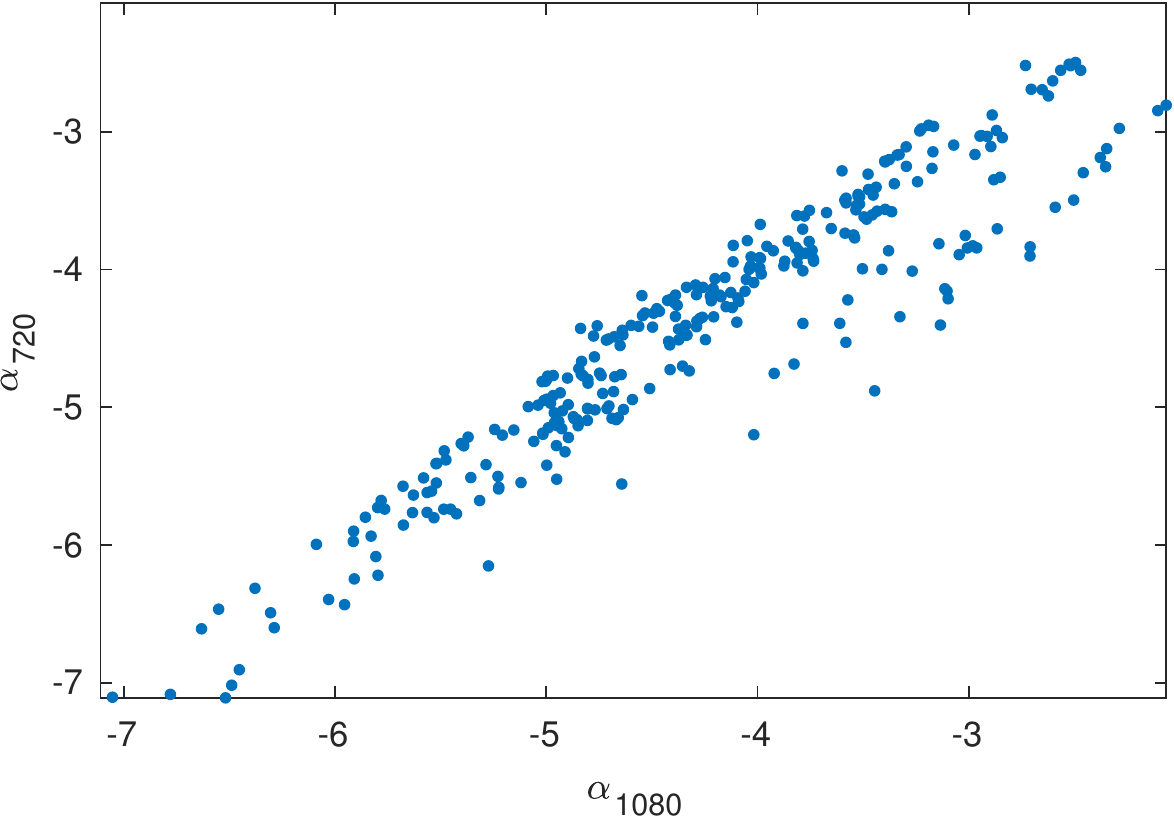}
			\footnotesize{(c) $\alpha_{720p}$ vs $\alpha_{1080p}$ (LCC: 0.9420, SROCC: 0.9412).}
		\end{minipage}
		\hfill
		\begin{minipage}{.48\linewidth}
			\centering
			\includegraphics[width=\linewidth]{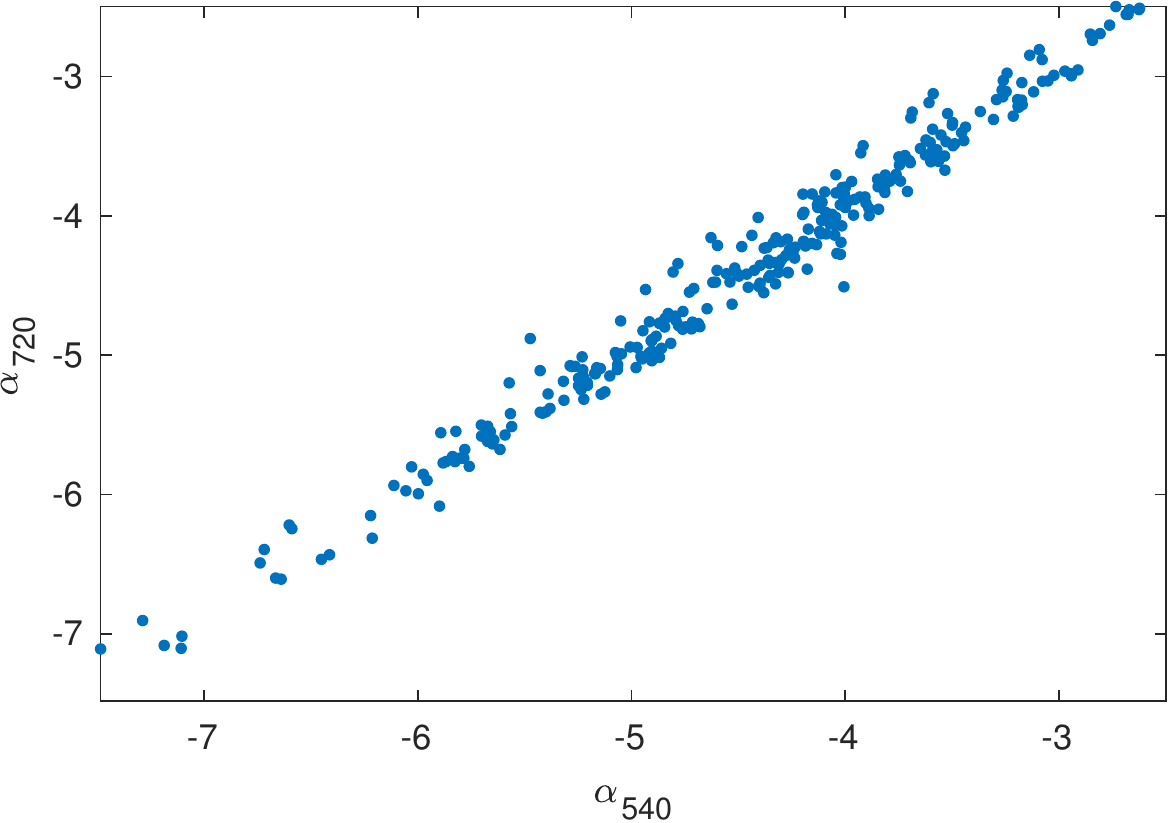}
			\footnotesize{(d) $\alpha_{720p}$ vs $\alpha_{540p}$, (LCC: 0.9890, SROCC: 0.9858).}
		\end{minipage}
		\caption{Exploring the QP-$\log$(R) model parameters across resolutions.}
		\label{fig:QP-logRate}
	\end{figure}

	In the construction of the bitrate ladder after predicting the cross-over points, we need to know which resolution to pick for each rung of the ladder and which QP corresponds to the respective bitrate. In order to predict the QP that corresponds to each bitrate ladder rung, we explored the QP-$\log$(R) relation. An example of QP-$\log$(R) points across three resolutions is illustrated in Fig.~\ref{fig:QP-logRate}(a). As indicated, we confirmed that there is a strong linear correlation with an average LCC equal to: -0.9891 for 2160p, -0.9931 for 1080p, -0.9955 for 720p, and -0.9952 for 540p. Thus, by defining a set of $\log$(R)-QP linear equations (one per resolution), we can estimate the $\widehat{P}_{L}$. Put formally:  
	\begin{equation}
	\widetilde{QP}_{s}=\alpha_{s} \log(R) + \beta_{s} \; ,
	\label{eq: QP(R)}
	\end{equation}
	where $\alpha_{s}, \; \beta_{s} \in \mathds{R}$ with $s \in \mathcal{S}$. So, each $\widehat{P}_{L,i}$ at $R_{L,i}$ can be estimated by this equation. We explored whether the $\alpha_{s},\beta_{s}$ parameters for the set of resolutions are correlated and noticed that there is a strong correlation between the $\alpha$ values, particularly in the lower resolutions, between 720p and 540p with LCC equal to 0.9890 and SROCC equal to 0.9858. This can be observed in Fig.~\ref{fig:QP-logRate}(b)-(d). Moreover, by fitting a first order polynomial, we noticed that $\alpha_{720p} \approx \alpha_{540p}$. This means that only one set of $(QP,\log(R))$ values is required to determine the $\beta_{540p}$ model parameter for 540p resolution. The same cannot be applied on the higher resolutions, as the deviation of the estimated $\alpha$ value is significant.
	
	In the considered example, where $|\mathcal{S}|=4$, we need to perform two encodes in one of the four resolutions and one encode for the remainder in order to fully define the RQ relations across resolutions. Naturally, it is more efficient to perform the two encodes at lower resolutions.

	\subsection{Compared Methods}
	Ideally, we would validate our proposed method against those state-of-the-art technologies described in Section~\ref{sec: RelatedWork}. However, as those are proprietary, with no publicly available implementations, we instead have benchmarked using the following methods. 
	\begin{itemize}
		\item \textit{Reference Ladder (RL)}: This exhaustive search approach was used to construct our reference Pareto surface, as described in Section~\ref{ssec: ConstrTheorCH} and Algorithm~\ref{alg: algorithm1}. To summarise, we encoded each sequence at different resolutions for a wide range of QP values, computed the cross-over QP's, constructed the PF, and the bitrate ladders. This method creates the optimal bitrate ladders and requires the highest number of encodings.
		\item \textit{Interpolation-based Ladder (IL)}: This method is based on encoding using only a subset of QP values per resolution. Specifically, after encoding using a subset of QPs per resolution, we then use a piece-wise cubic Hermite interpolation~\cite{pchip} to find the RQ coordinates for the interim QP values. Based on these encodings and and the interpolated RQs, the PF is extracted as in the RL method explained above. This method produces a suboptimal solution, whose accuracy depends on the number of encodes performed per resolution. The added benefit of this method is that it significantly reduces the number of encodings required compared to the RL.
		\item \textit{Feature-based Predicted Ladder (FL)}: This is the proposed method described earlier in Algorithm~\ref{alg: algorithm2}, where spatio-temporal features are extracted first to predict the RQ cross-over points that are on the PF. Then encodings at the cross-over QPs are used to define the bitrates, where resolution switches, and to estimate the parameters of Eq.~(\ref{eq: QP(R)}). After the estimation of the parameters, the equations are utilised along with the switching bitrates to estimate the QP values and the resolution for the bitrate ladder rungs.
		\item \textit{Hybrid Ladder (HL)}: This method combines the best performing method per content, either FL or IL. A method selection module is introduced after the spatio-temporal feature extraction. Using the extracted spatio-temporal features, a classifier selects for each input sequence which method, IL or FL, is expected to more accurately estimate the bitrate ladder. 
	\end{itemize}

	\subsection{Bitrate Ladder Prediction Results}
	
	As described in Algorithm~\ref{alg: algorithm2}, after predicting the cross-over QPs, we perform encodings at the defined cross-over points in order to estimate the two parameters of Eq.~(\ref{eq: QP(R)}) at each resolution. After defining the parameters, the bitrate ladders for the considered fitted models are constructed following the approach described in Section~\ref{ssec: BuildLadder}. In order to assess the predicted bitrate ladder against the two benchmarks, we computed the BD metrics~\cite{Bjontegaard}, BD-Rate and BD-PSNR per sequence.
	For the comparison of the methods, we  report the mean values and the mean absolute deviation of both metrics in Table~\ref{tab: predPF}. As an additional measure of optimality, this Table also reports the average percentage of the predicted RQ points that belong to the PF (PF-hits). We selected the mean absolute deviation (mad) instead of standard deviation because, as easily observed in the histograms of Fig.~\ref{fig: Histograms}, the distributions are not normal. The distributions are skewed due to the fact that the BD metrics against the RL bitrate ladders that are constructed from points on the PF. The juxtaposed methods are potentially composed by a mixture of points that either belong to the PF or to a suboptimal set of points. Moreover, in Fig.~\ref{fig: ExamplesSeqsLadders}, we provide examples of predicted ladders using all methods for different sequences.

	\subsubsection{IL Method}
\begin{figure}[!t]
		\begin{minipage}{.49\linewidth}
			\centering
			\includegraphics[width=\linewidth]{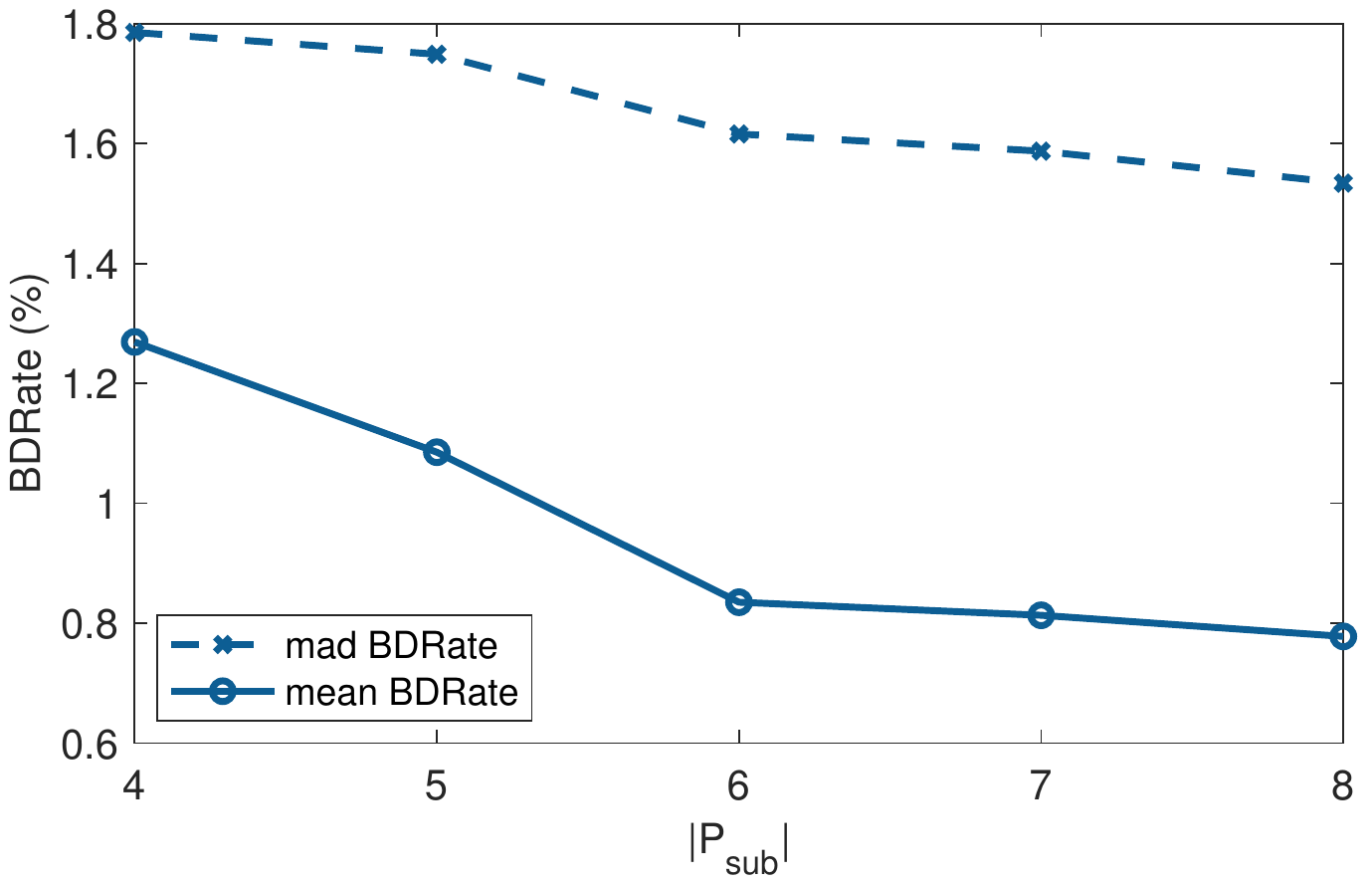}
			\vspace{2px}
			\footnotesize{(a) BD-Rate.}
		\end{minipage}
		\begin{minipage}{.49\linewidth}
			\centering
			\includegraphics[width=\linewidth]{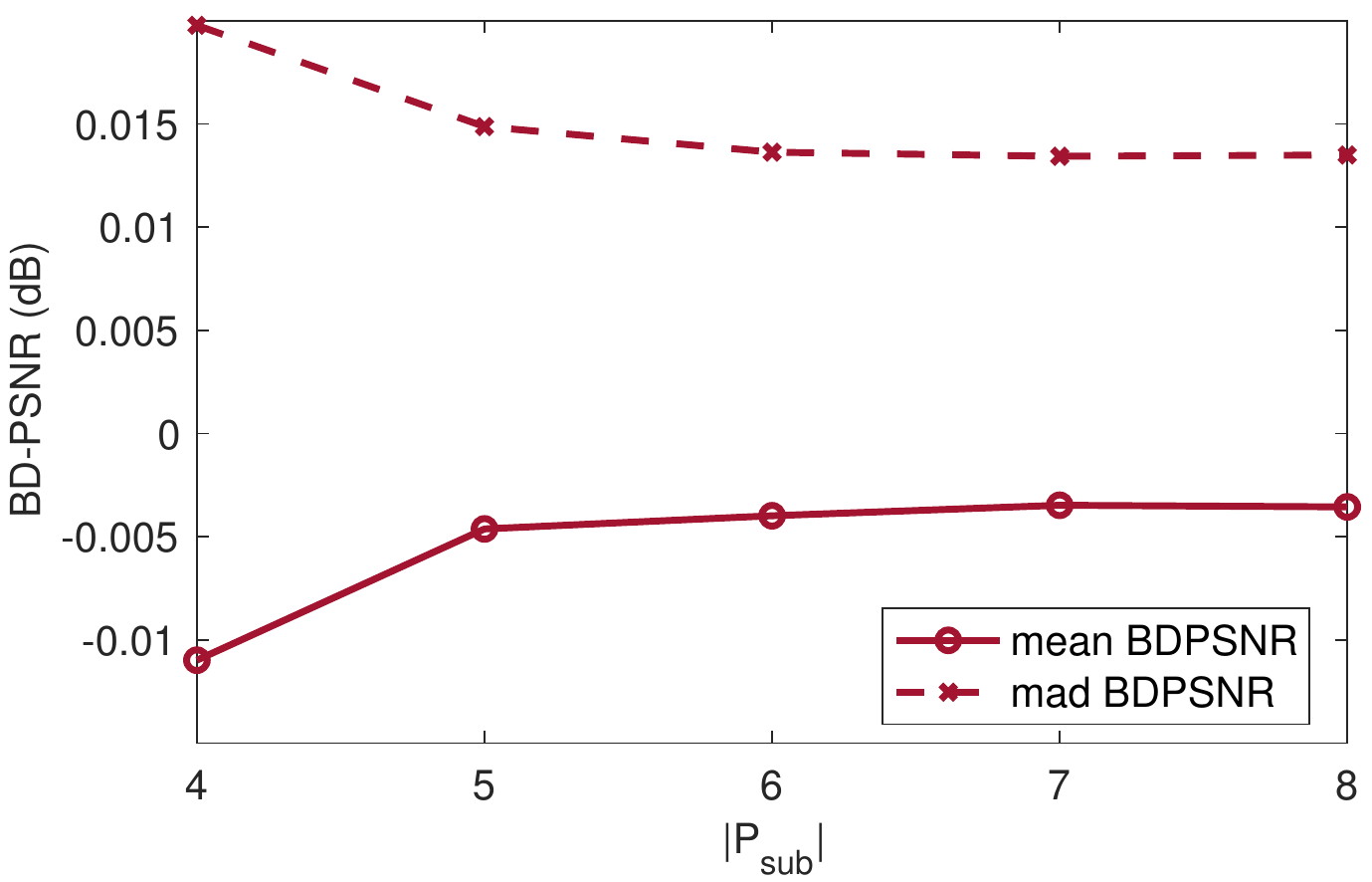}
			\vspace{2px}
			\footnotesize{(b) BD-PSNR.}
		\end{minipage}
		\caption{Mean BDRate and confidence intervals of the IL over the RL against the number of encodes required.}
		\label{fig: ILvsNoEncodes}
	\end{figure}

	We first investigated the accuracy of the IL method by computing the BD metrics for varying QP samples $|\mathcal{P}_{sub}|$ per resolution, namely from 4 to 8. Figure~\ref{fig: ILvsNoEncodes} plots the mean BD-Rate$_\textrm{P}$ and BD-PSNR with their mean absolute deviation for the different $|\mathcal{P}_{sub}|$ encodes per resolution. As can be seen from this figure, 
	as the number of encodes increases the mean BD-Rate drops resulting to a very good approximation of the RL, as also verified by the results reported in Table~\ref{tab: predPF}. The variations in the results are mainly attributed to the sensitivity of the interpolation method and the different number of steps. The BD statistics start converging for $|\mathcal{P}_{sub}|\geq6$ with the best results achieved for$|\mathcal{P}_{sub}|=7$. From this point onward, we will be using this setting to compare with the other methods. For this case, as reported in Table~\ref{tab: predPF} and shown in Fig.~\ref{fig: Histograms}, the mean BD-Rate is 0.80\% with a mean absolute deviation of 1.71\%, while the BD-PSNR is -0.004dB with a mean absolute deviation of 0.001dB. Besides this, the IL method results in ladders composed with 87.5\% RQ points from the PF. This is a strong indication of the optimality of the predicted ladder.
	
	The histograms of BD-Rate and BD-PSNR for 7 QP samples per resolution are plotted in Fig.~\ref{fig: Histograms}(a)-(b). As can be seen,  the distributions are tightly clustered around the mean values, while it can also be observed that, for a small number of sequences, the |BD-Rate|$>$5\%.

	\subsubsection{FL Method}
	For the FL method, we first performed a number of initial encodes required to determine the parameters of Eqs.(\ref{eq: QP(R)}) that will lead to the QP that corresponds to the rung bitrate. The predicted cross-over QPs, $\{ \widehat{QP}^{high}_{2160}$, $\widehat{QP}^{low}_{1080}$, $\widehat{QP}^{high}_{1080}$, $\widehat{QP}^{low}_{720}$, $\widehat{QP}^{high}_{720}$, $\widehat{QP}^{low}_{544} \}$, are used for the initial encodes. With the RQ points at $\widehat{QP}^{low}_{1080}, \widehat{QP}^{high}_{1080}$, the $\alpha_{1080p},\beta_{1080p}$ are defined. The RQ points resulting from the $\widehat{QP}^{low}_{720}, \widehat{QP}^{high}_{720}$ encodes are utilised to determine the $\alpha_{720p},\beta_{720p}$ parameters and the $\beta_{540p}$ parameter, as well. Additionally to the above six initial encodes, one more encode for the 2160p resolution is used. The QP value selection for the extra encode is decided based on $\widehat{QP}^{high}_{2160}$ value: if it is towards the lower end a higher QP is selected, and vice versa. The additional encode in 2160p helps to improve the predictions towards the higher bitrates because, as shown in Fig.~\ref{fig:QP-logRate}, the $\alpha_{2160p}, \alpha_{1080p}$ values deviate. 
	Also, after the bitrate ladder construction with the FL method, we performed a monotonicity and concavity check, as indicated in Step 5 of Algorithm~\ref{alg: algorithm2}. According to this, if non-monotonic points are detected, we sort them. Also, if a bitrate ladder point results in a concave RQ curve, we remove this point, as this most likely is a suboptimal point.
	
\begin{figure}[!h]
		\begin{minipage}{.49\linewidth}
			\centering
			\includegraphics[width=\linewidth]{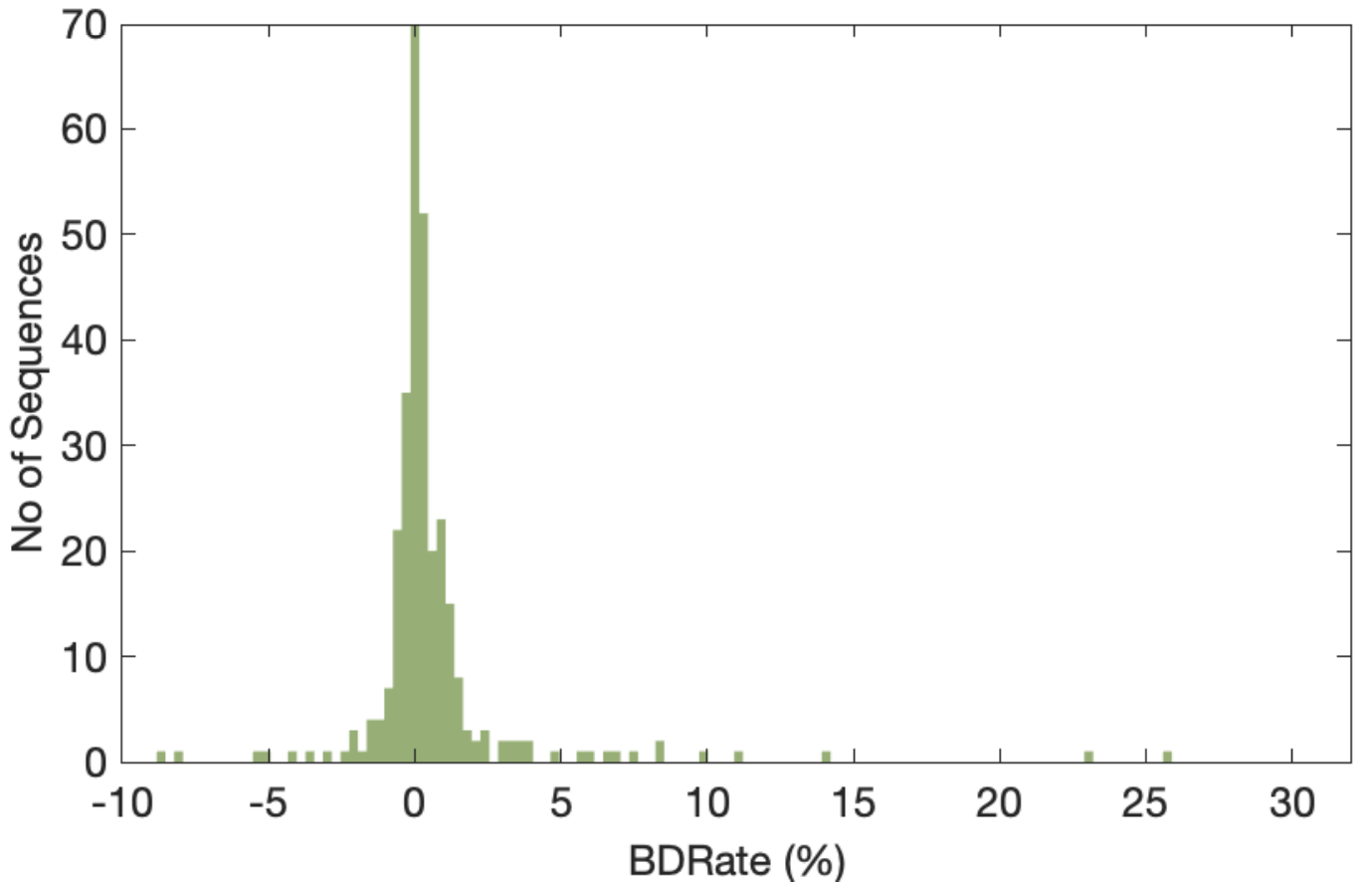}
			\vspace{2px}
			\footnotesize{(a) BD-Rate for IL.}
		\end{minipage}
		\begin{minipage}{.49\linewidth}
			\centering
			\includegraphics[width=\linewidth]{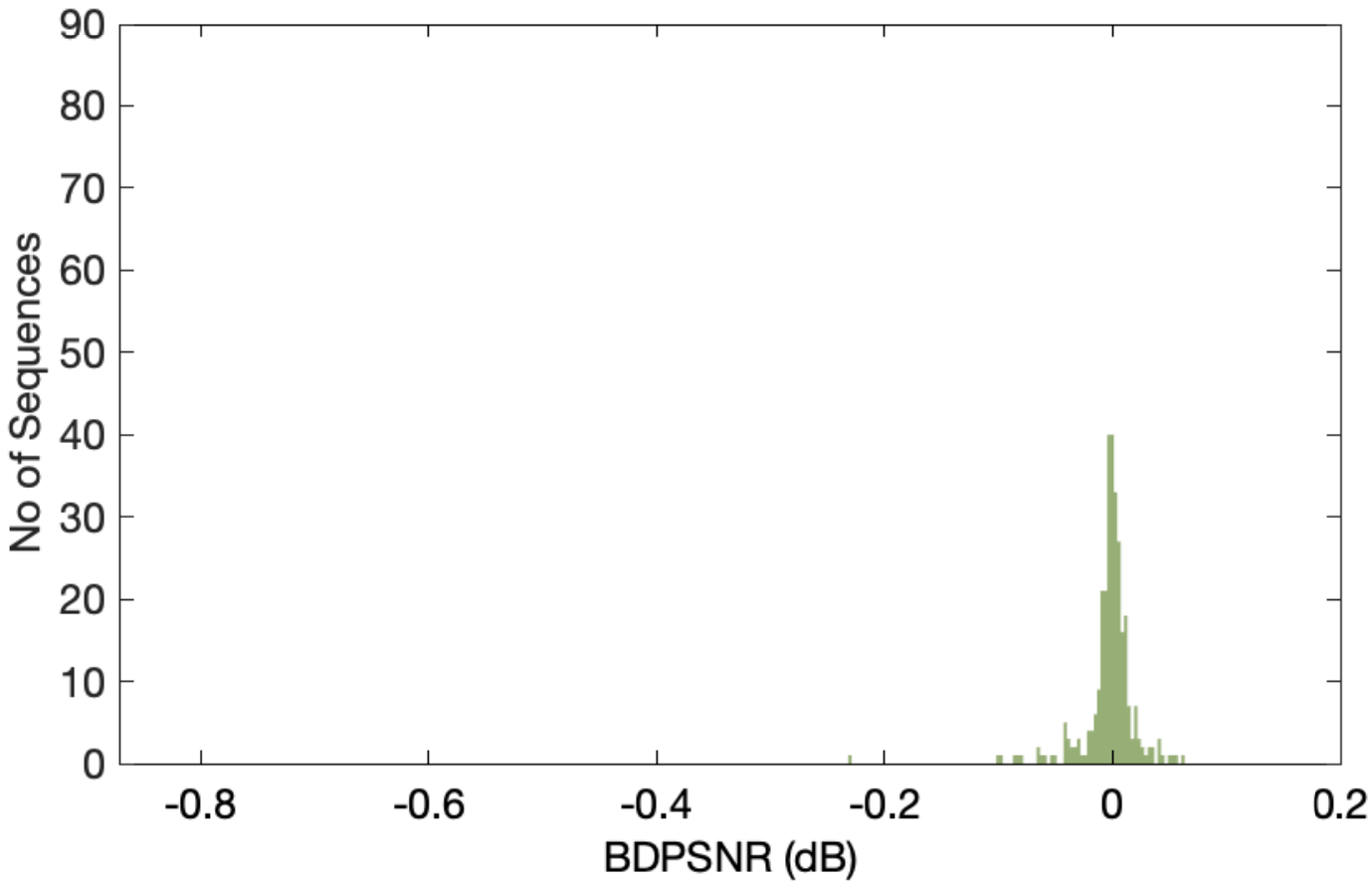}
			\vspace{2px}
			\footnotesize{(b) BD-PSNR for IL.}
		\end{minipage}
		\begin{minipage}{.49\linewidth}
			\centering
			\includegraphics[width=\linewidth]{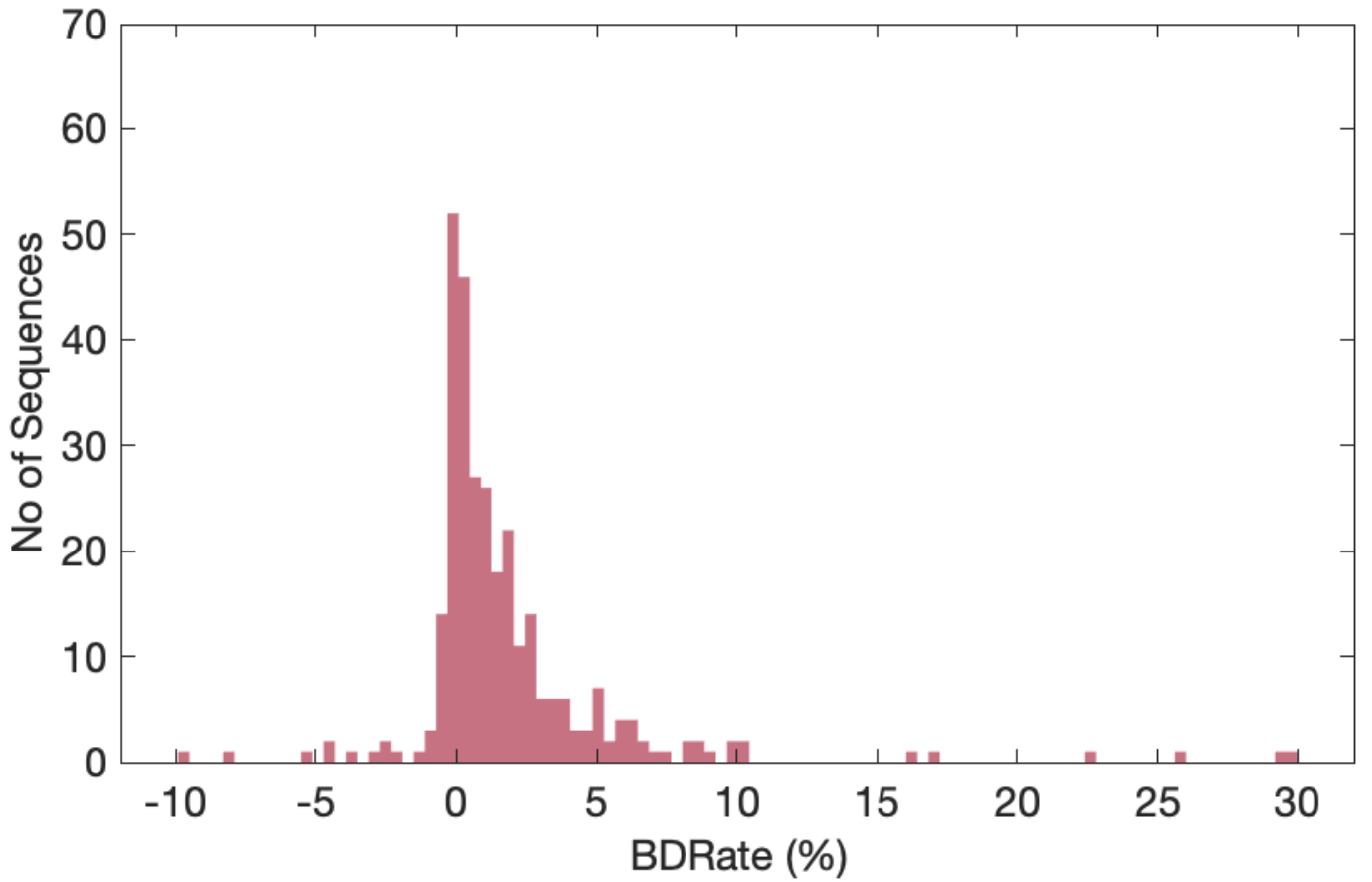}
			\vspace{2px}
			\footnotesize{(c) BD-Rate for FL.}
		\end{minipage}
		\begin{minipage}{.49\linewidth}
			\centering
			\includegraphics[width=\linewidth]{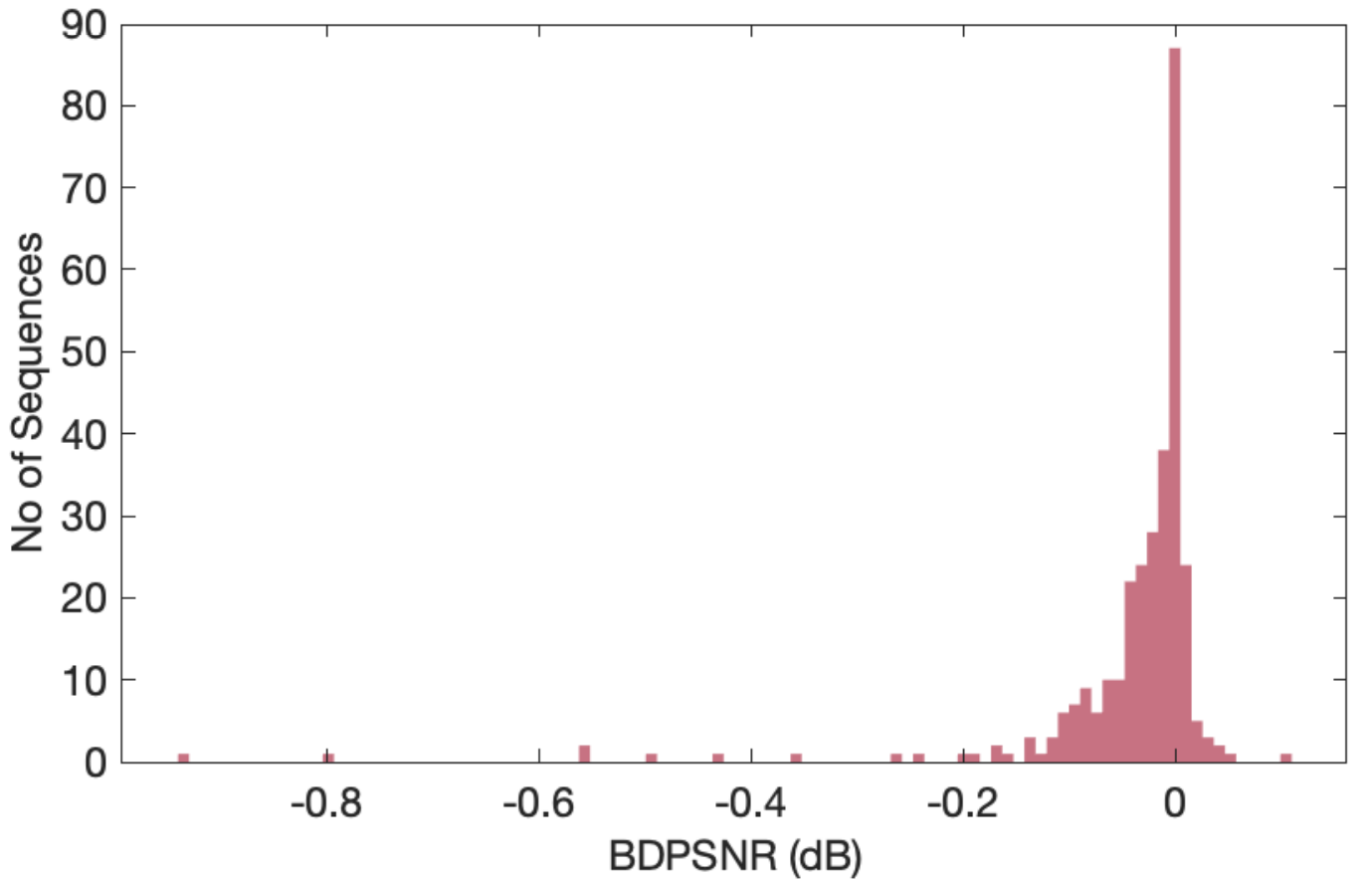}
			\vspace{2px}
			\footnotesize{(d) BD-PSNR for FL.}
		\end{minipage}
		\begin{minipage}{.49\linewidth}
			\centering
			\includegraphics[width=\linewidth]{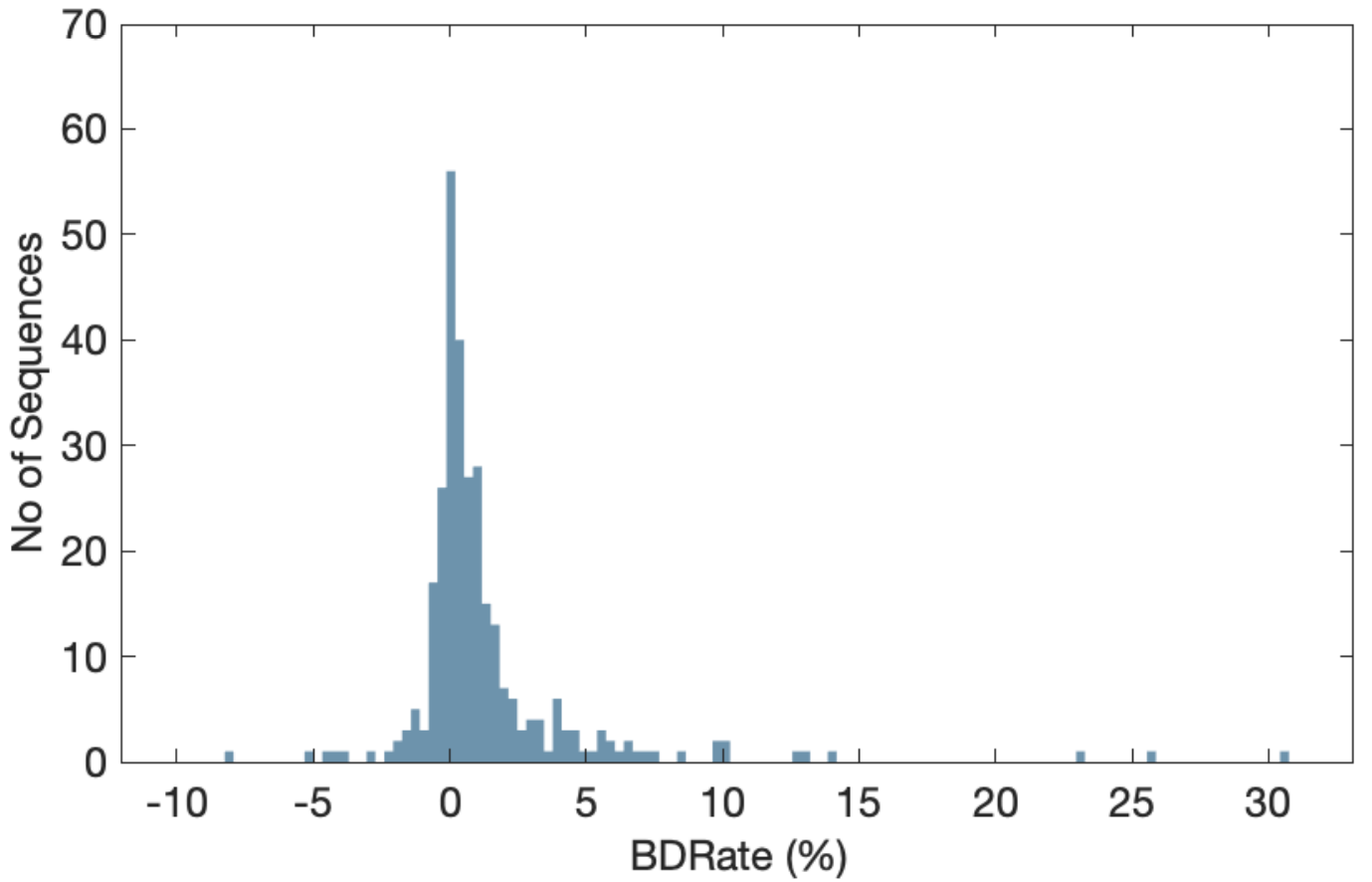}
			\vspace{2px}
			\footnotesize{(e) BD-Rate for HL.}
		\end{minipage}
		\begin{minipage}{.49\linewidth}
			\centering
			\includegraphics[width=\linewidth]{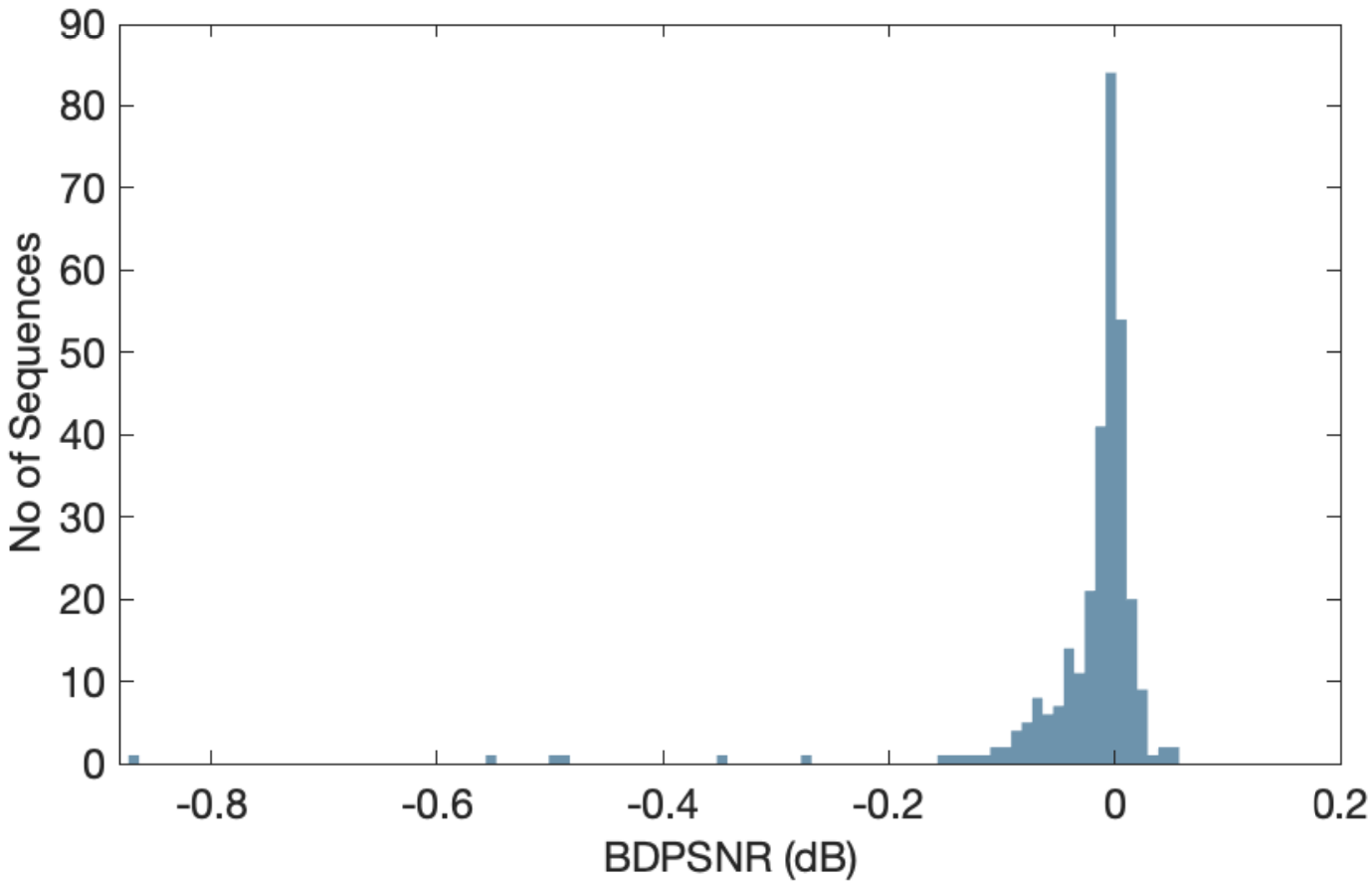}
			\vspace{2px}
			\footnotesize{(f) BD-PSNR for HL.}
		\end{minipage}
		\caption{BD metric histograms for the compared methods in Table~\ref{tab: predPF}.}
		\label{fig: Histograms}
	\end{figure}

	Inspecting the histograms of FL BD-Rate and BD-PSNR in Fig.~\ref{fig: Histograms}~(c)-(d) we observe that, compared to IL, these distributions have heavier tails, which is verified by the higher mean absolute deviation values.
	Table~\ref{tab: predPF} reports the mean and mean absolute deviation of the BD metrics of the proposed FL method. The mean BD-Rate is 0.98\% higher compared to IL, while the mean absolute deviation is increased by 0.5\%. Despite these increased figures, the PF-hits percentage remains high, over 80\%.
	
	\begin{table}[!ht]
		\caption{BD metrics for the predicted ladders with the proposed methods and percentage of points on the PF.}
		\centering
		\footnotesize
		\begin{tabular}{l|r|r|r}
			\toprule
			Method & [mean,mad]  BD-Rate & [mean,mad] BD-PSNR & PF-hits\\
			\midrule
			IL vs RL & 0.80\%, 1.71\% & -0.004dB, 0.001dB & 87.50\%\\
			FL vs RL&  1.78\%, 2.27\% & -0.04dB, 0.05dB & 80.48\%\\
			HL vs RL& 1.26\%, 1.91\% & -0.02dB, 0.04dB & 83.86\% \\
			\bottomrule
		\end{tabular}
		\label{tab: predPF}
	\end{table}

	\subsubsection{HL Method}
\begin{figure}[!h]
		\centering
		\includegraphics[width=.6\linewidth]{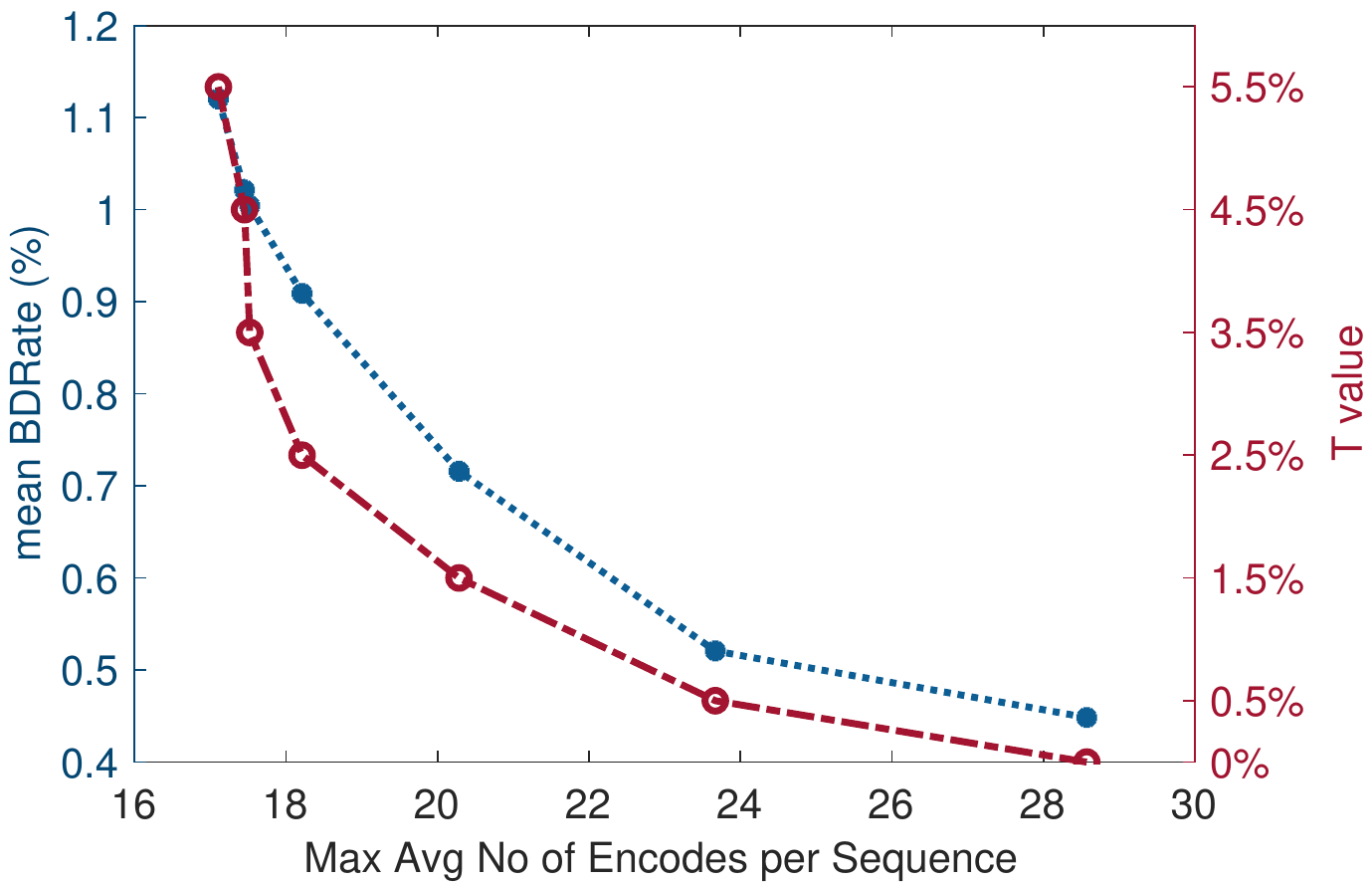}
		\vspace{2px}
		\caption{Effect of the HL threshold for the expected accuracy of predictions and the number of encodes.}
		\label{fig: HLvsT}
	\end{figure}

	We investigated the effectiveness of a hybrid approach that combines the IL and FL methods. In many cases, IL and FL construct almost identical ladders leading to a very similar BD-Rate when compared to the RL. Therefore, the rule that we apply is that the IL method should be chosen only for those cases, where it will improve the FL BD-Rate at least by a threshold $T$. We explored the impact of this threshold by selecting a set of values, $T \in \{0\%, 0.5\%, 1.5\%, 2.5\%, 3.5\%, 4.5\%, 5.5\%\}$ and by estimating the best BDRate that could be achieved if the selection of the method was performed with a 100\% accuracy. In Fig.~\ref{fig: HLvsT}, we illustrate the effect of this threshold on the mean BD-Rate and the maximum expected average number of encodes per sequence. As clearly illustrated, the increase of the $T$ value results in an increase of the mean BD-Rate while decreasing the average required number of encodings. From the BD-Rate to the average number of encodes tradeoff, we selected $T=0.5\%$ as it appears as an optimal threshold.
	
	After defining the threshold, we proceeded to the method selection step. To this end, before predicting  the cross-over QPs, we implemented a binary classifier. If the IL method is selected for a sequence, then we proceed as described above. If the FL method is selected, we proceed with the cross-over QPs prediction to apply the FL method. The classifier was built using Ensemble Trees utilizing the set of spatio-temporal features F1-F20 and a ten-fold cross-validation was performed to avoid overfitting. The resulting accuracy of classification was 68\%.
	As a result, for 70\% of the sequences the FL method was selected, while for 30\% of the sequences the IL method was predicted to be more accurate. 
	
	The resulting BD statistics from this hybrid method, after the method selection process, are illustrated in Fig.~\ref{fig: Histograms}(e)-(f). Although the classification accuracy was not very high, it is evident that these results show significant improvements compared to the FL results with statistics closer to those from IL in (a) and (b). The mean BD-Rate value is .52\% lower than that of FL and 0.46\% higher of IL, while the mean absolute deviation is almost equal to of IL. What is further significant is that the PF-hits value for HL slightly drops (only 3.64\%), indicating the optimality of the predicted ladders.

	\subsection{Relative Complexity}
	The proposed method FL and its hybrid combination with IL, HL, both offer a close-to-optimal prediction of the bitrate ladder, while achieving a significant reduction in the number of encodings required to build the ladder for any new video sequence. In Table~\ref{tab:complex}, the numbers of encodings required for the compared methods are reported. As explained earlier, the exhaustive-search method produces the optimal PF but this comes at the cost of $|\mathcal{S}|\times|\mathcal{P}|$ encodings needed to derive it. In the considered test case, the number of encodings needed is 132. 
	
	For the IL method, based on the results of Fig.~\ref{fig: ILvsNoEncodes}, seven encodes per resolution were selected. This means that a total number of $|\mathcal{S}|\times7=28$ encodes is required per sequence in order to find the rate points where resolution switches and build the PF. In this case, an additional number of encodes $E_req$, with $E_{req}\in\{0,1, \ldots,|\mathcal{L}|\}$, is required to build the bitrate ladder. This varies for each sequence as it depends from the length of its ladder. The average recorded number of encodes for the considered dataset was 35.21. This method brings a significant reduction of 71.60\% compared to the RL method in the presented test case. 
	
	The FL method requires initially only $2 \times |\mathcal{S}|-1=7$ encodes to define the rate points, where resolution switches, and to compute the parameters of Eq.~(\ref{eq: QP(R)}) for each sequence. Then, similarly to IL, $E_{req}$ number of encodes are required to hit the target bitrates at each ladder rung. In the presented test case, 13.57 encodings were required on average for FL. Clearly, FL outperforms both RL and IL approaches in terms of number of encodes, requiring from 89.06\% fewer encodings compared to the RL method and about 61.46\% less encodes compared to IL. 
	
	The hybrid method HL offers an important improvement in terms of required number of encodings per sequence while producing a very close to optimal ladder. This depends of course on the number of sequences and how often IL or FL is invoked. For the presented test case, where for 70\% of the sequences the FL method was selected for the ladder construction, an average of 20.05 encodes was performed resulting to a 83.83\% reduction compared to RL and to a 43.06\% reduction compared to IL. Compared to pure FL, HL performed on average 6.48 more encodes.
	
	Although FL achieves an important reduction in the number of encodings required, it also introduces an overhead associated with the computation of the extracted features and with the cross-over QP prediction. The cost of the  the construction of the bitrate ladder is thus almost identical for IL and FL methods. The ratio of the average feature extraction time for a sequence at 2160p resolution to the average 2160p encoding time for a sequence at one QP is 0.18\footnote{The feature extraction is implemented in Matlab, while for the encodings the HM reference software was used.}. The cross-over QP prediction time is negligible compared to encoding time. Considering this, FL's complexity is still significantly lower from that of the IL method. 
	
	\begin{table*}[h]
		\begin{center}
			\caption{Comparison of the number of encodings required per method for each sequence.} 
			\label{tab:complex}
			\footnotesize
			\begin{tabular}{c|c|c|c}
				\toprule
				\textbf{Method} &  \textbf{\# Encodings}& \textbf{Test Case} ($|\mathcal{S}|=4$)& \textbf{ Average \# Encodings} \\
				\midrule
				RL& $|\mathcal{S}|\times|\mathcal{P}|$ & 4$\times$31 & 124\\
				\midrule
				IL  & $|\mathcal{S}|\times|\mathcal{P}_{sub}|+E_{req}$, with $E_{req}\in\{0,1, \ldots,|\mathcal{L}|\}$ & 28+(from 0 up to $|\mathcal{L}|$)& 35.21\\ 
				\midrule
				FL  & $(2 \times |\mathcal{S}|-1) +E_{req}$ & 7+(from 0 up to $|\mathcal{L}|$)& 13.57\\
				\midrule
				HL  & $|\mathcal{S}| \times |\mathcal{P}_{sub}| \; \parallel \; (2 \times |\mathcal{S}|-1 ) +E_{req}$ & (28||7)+(from 0 up to $|\mathcal{L}|$)& 20.05\\
				\bottomrule
			\end{tabular}
		\end{center}
	\end{table*}
	
	\subsection{Discussion}
	From the results discussed above, the FL and HL methods offer a significant reduction in the required number of encodes compared to RL or IL for only a small BDRate cost. FL and HL solutions are close to optimal with over 80\% of the bitrate ladder points produced belonging to the PF. Generally, IL, FL, and HL offer very similar solutions and produce bitrate ladders with a high percentage of points on the PF. Suboptimal points of lower or higher target bitrate, could potentially be improved with an additional round of encodes at an incremental QP to result in an attempt to hit the bitrate closer to the target.
	
\begin{figure}[!ht]
		\begin{minipage}{.49\linewidth}
			\centering
			\includegraphics[width=\linewidth]{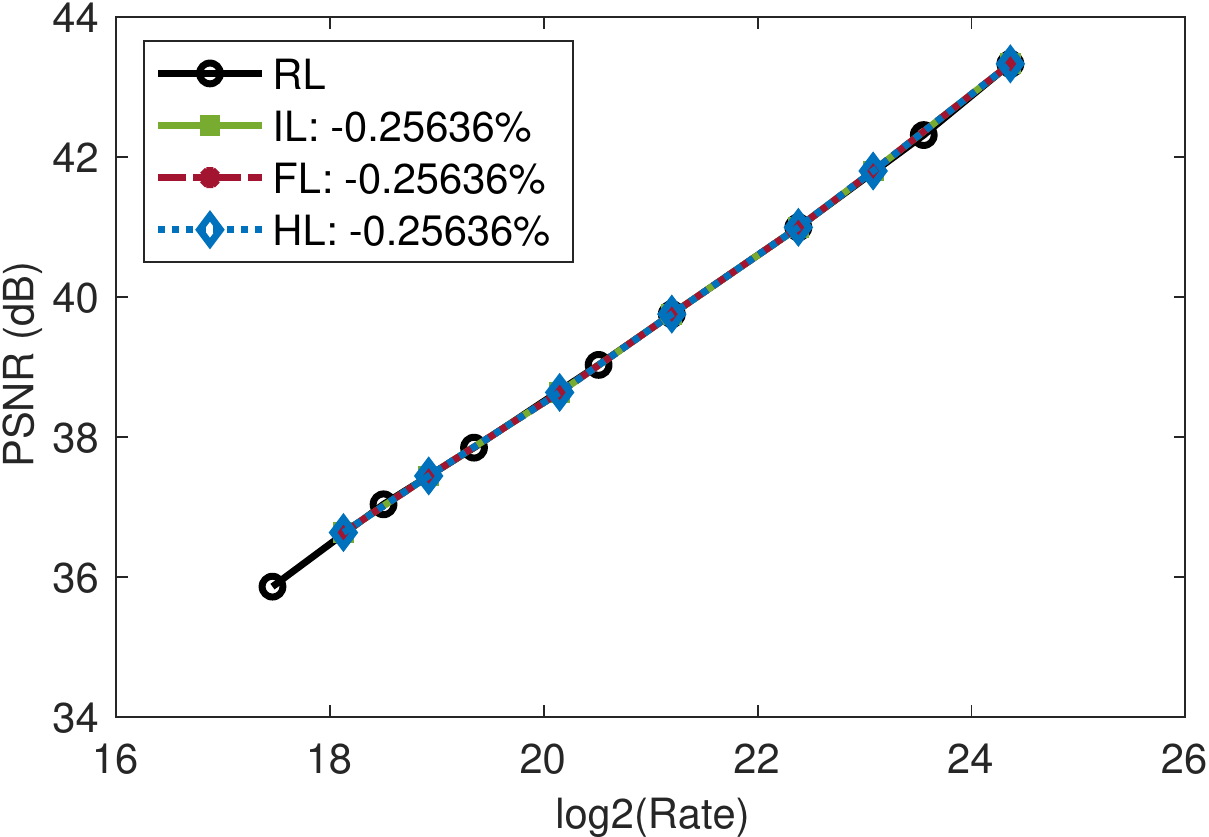}
			\vspace{2px}
			\footnotesize{(a) Air-acrobatics.}
		\end{minipage}
		\hfill
		\begin{minipage}{.49\linewidth}
			\centering
			\includegraphics[width=\linewidth]{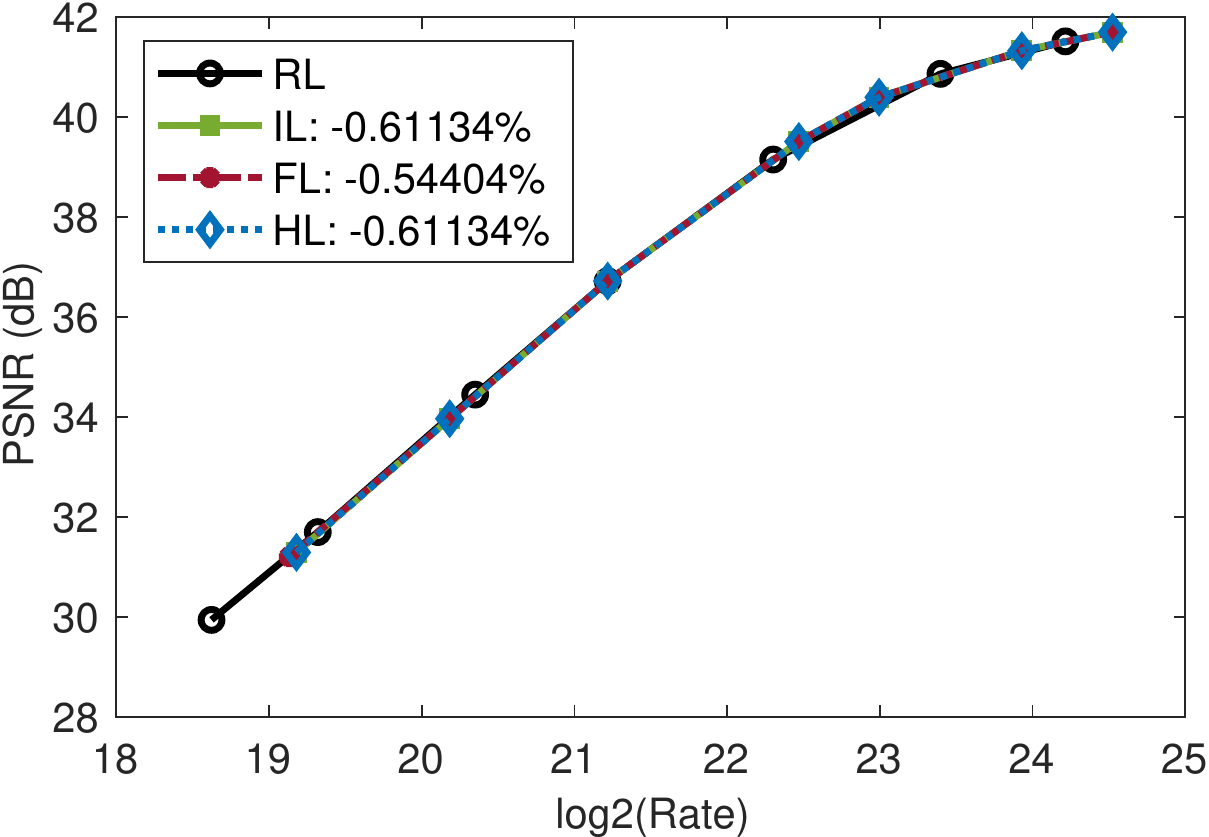}
			\vspace{2px}
			\footnotesize{(b) BoxingPractice.}
		\end{minipage}
		\begin{minipage}{.49\linewidth}
			\centering
			\includegraphics[width=\linewidth]{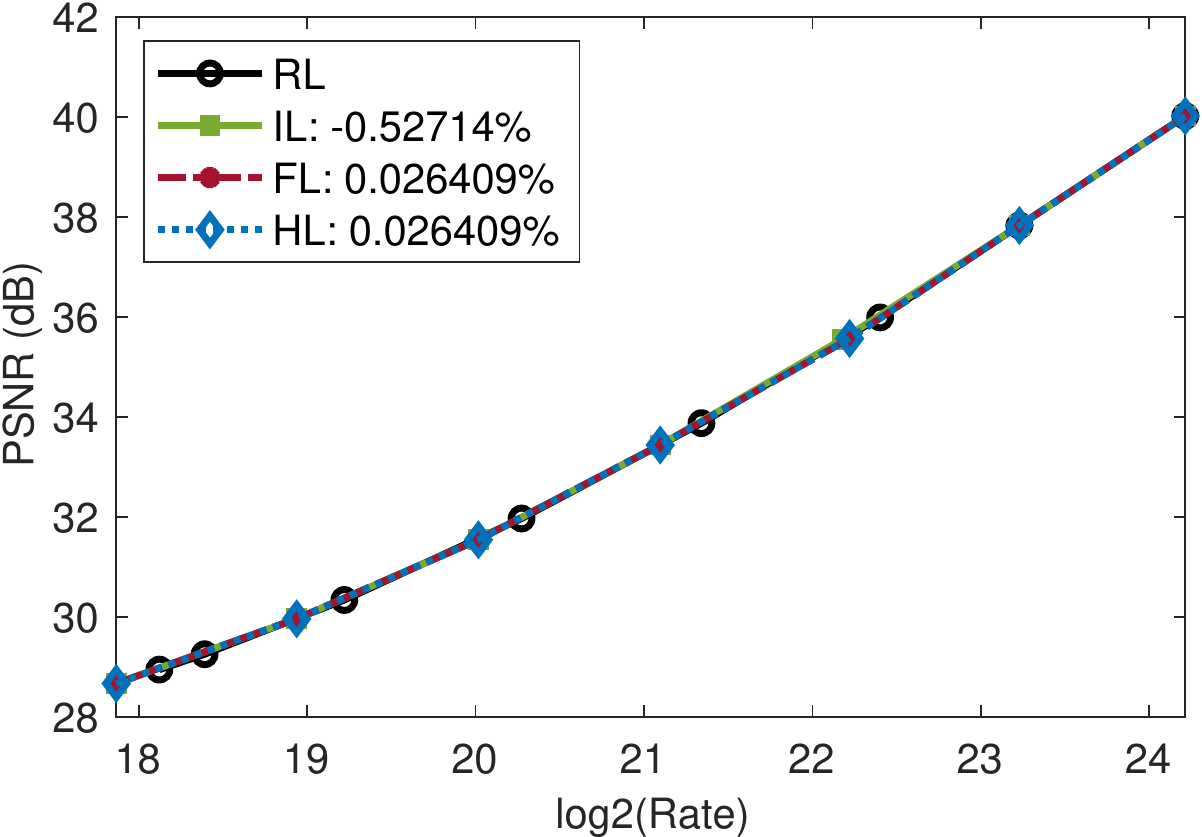}
			\vspace{2px}
			\footnotesize{(c) Coastguard.}
		\end{minipage}
		\hfill
		\begin{minipage}{.49\linewidth}
			\centering
			\includegraphics[width=\linewidth]{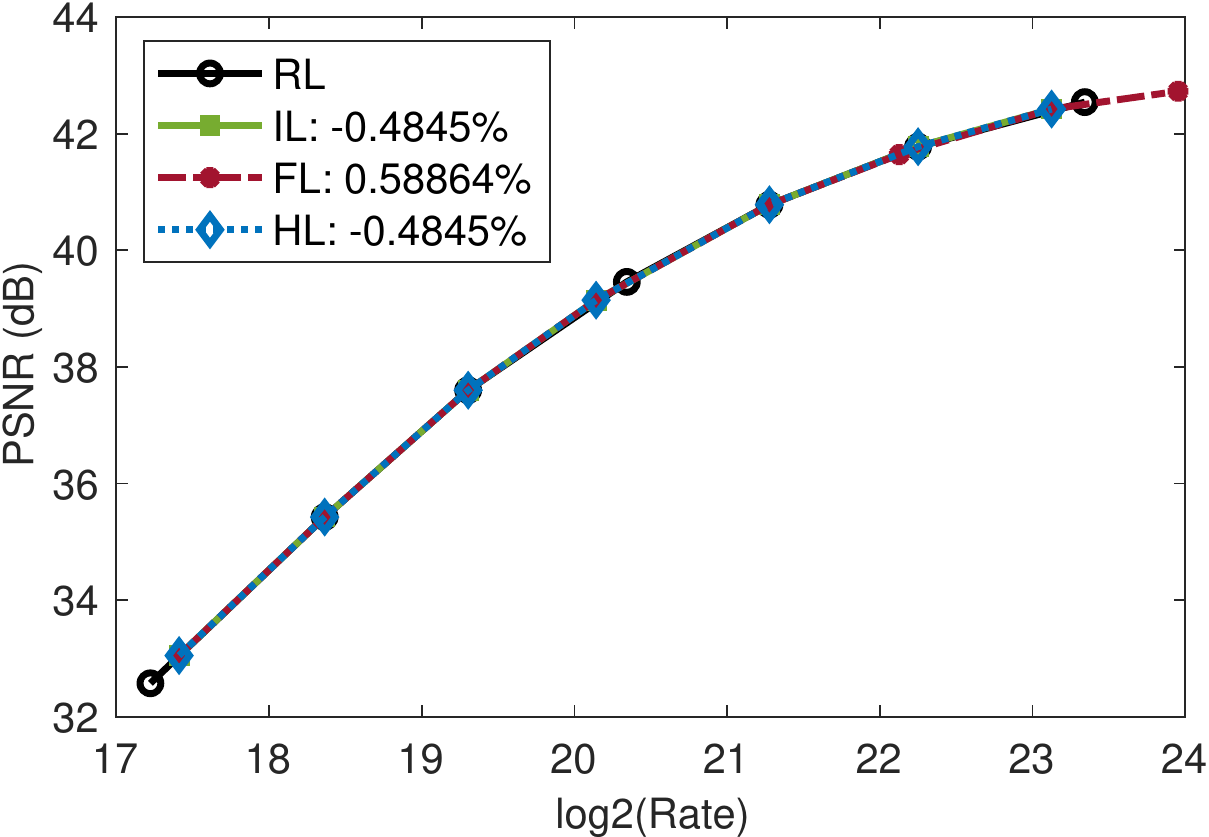}
			\vspace{2px}
			\footnotesize{(d) Crosswalk.}
		\end{minipage}
		\begin{minipage}{.49\linewidth}
			\centering
			\includegraphics[width=\linewidth]{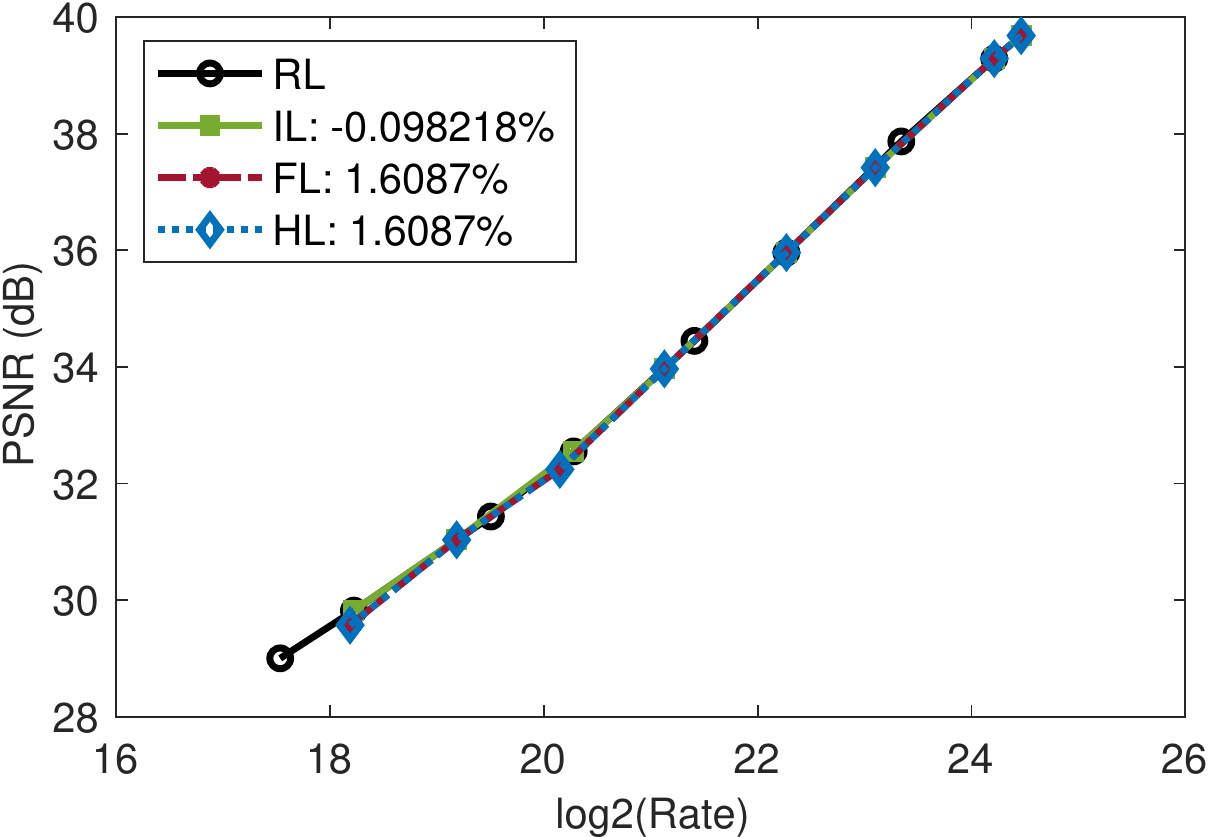}
			\vspace{2px}
			\footnotesize{(e) Treeshade.}
		\end{minipage}
		\hfill
		\begin{minipage}{.49\linewidth}
			\centering
			\includegraphics[width=\linewidth]{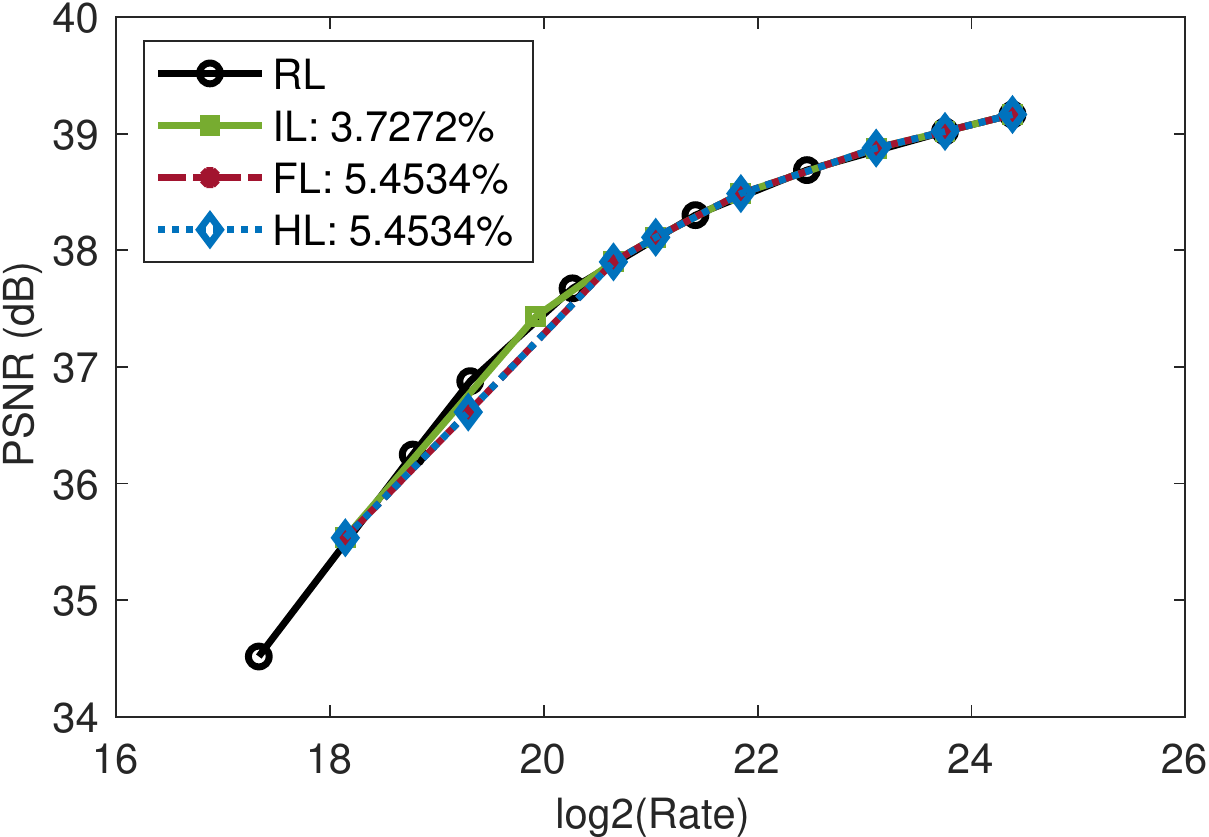}
			\vspace{2px}
			\footnotesize{(f) Raptors.}
		\end{minipage}
		\hfill
		\begin{minipage}{.49\linewidth}
			\centering
			\includegraphics[width=\linewidth]{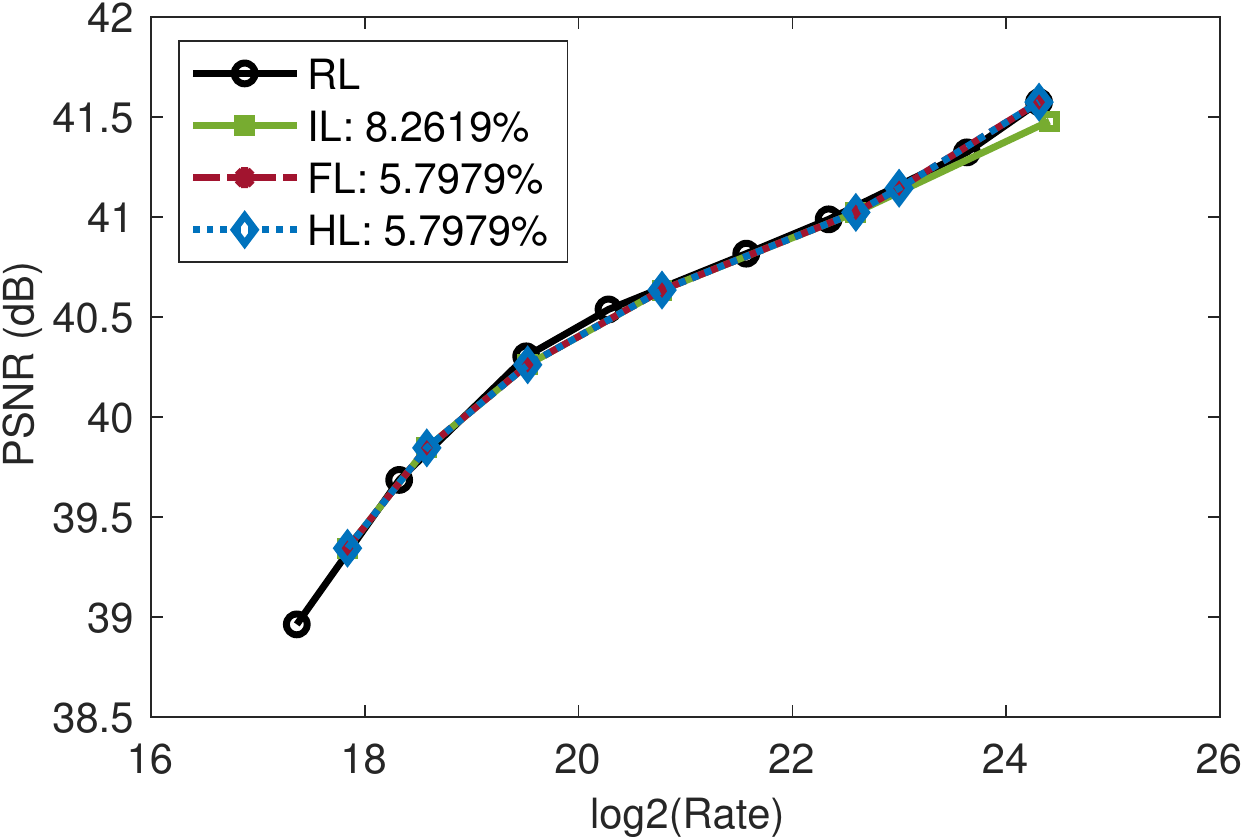}
			\vspace{2px}
			\footnotesize{(g) Skateboarding-scene8.}
		\end{minipage}
		\begin{minipage}{.49\linewidth}
			\centering
			\includegraphics[width=\linewidth]{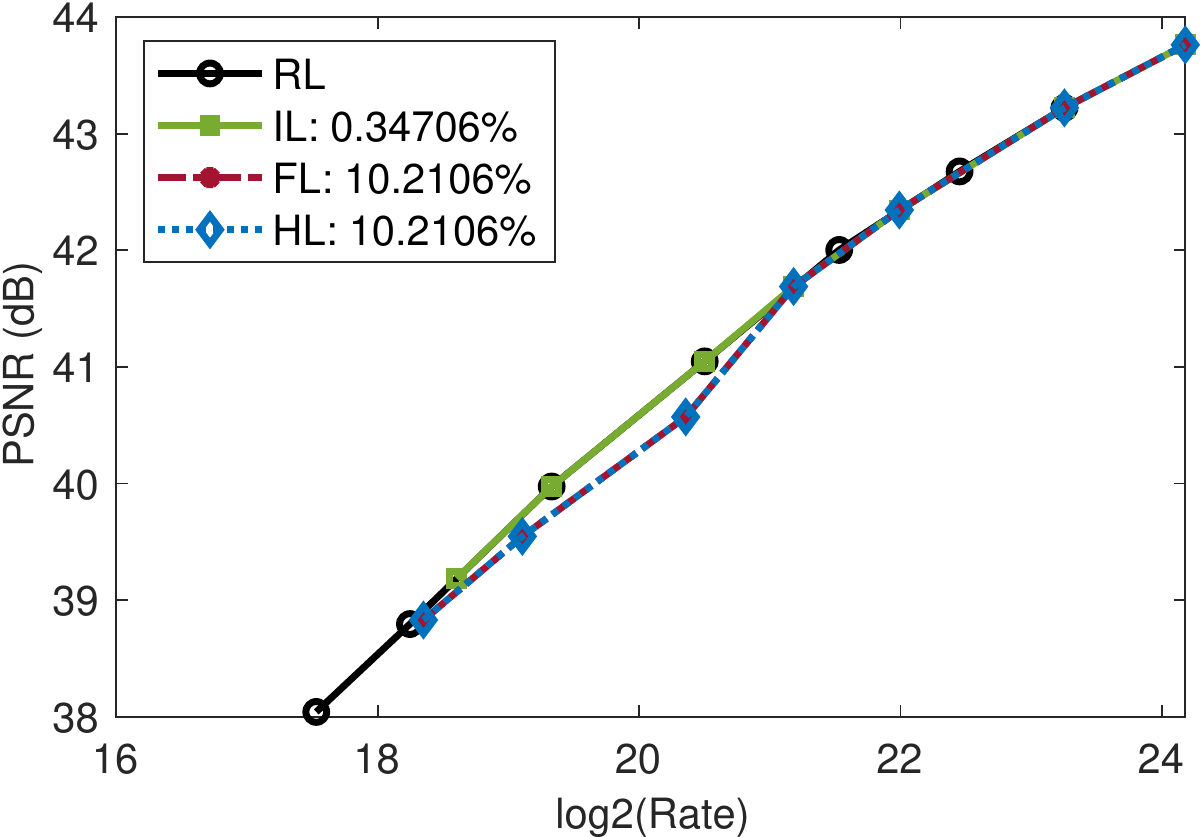}
			\vspace{2px}
			\footnotesize{(h) WindAndNature-scene2.}
		\end{minipage}
		\caption{Examples of predicted ladders for different sequences with the BD-Rate reported per tested method.}
		\label{fig: ExamplesSeqsLadders}
	\end{figure}

	Regarding the distributions of the BD statistics, we observed two kinds of outliers. On one hand, negative BD-Rate values are observed in all three methods against the expectation of only positive BD-Rate values. The negative BD-Rates are attributed to the fact that in many cases the bitrate ladder might be composed of points on the PF that are shifted towards higher bitrates and PSNR. Thus, in curved PSNR-$\log$(R) ladders those segments create a raised PF. An example of this is provide in Fig.~\ref{fig: ExamplesSeqsLadders}~(e). On the other hand, we noticed that for all tested methods, IL, FL, and HL, there are outliers of |BD-Rate|$>$5\% which generally is considered an important deviation from the reference. Examples of those are given in Fig.~\ref{fig: ExamplesSeqsLadders}~(f)-(h). These statistics are either caused by suboptimal points or by not matching all the ladder rungs. For example in Fig.~\ref{fig: ExamplesSeqsLadders}~(g), most of the points are not the PF for all methods (PF-hits: 2/7 for IL and 3/7 for FL). This specific sequence produces RQs with an unusual short range of PSNR, 39-41.5dB within the wide range of [500kbps,25Mbps]. This means that although the BD-Rate might be considerable the impact on the resulting quality is insignificant. Other outliers, such as Fig.~\ref{fig: ExamplesSeqsLadders}~(h), that are observed in the FL and HL method are attributed to incorrect predictions of the cross-over QPs that subsequently lead to an incorrect bitrate for the resolution switching. Thus, the predicted ladder is suboptimal.
	
	The other examples of RQs in Fig.~\ref{fig: ExamplesSeqsLadders}~(a)-(d), indicate cases of successful prediction of both IL and FL methods with a high PF-hits percentage (over 75\%). Also, they indicate cases for HL, when the correctly better method was selected through the classification step or cases where it failed. In most cases though the BD-Rate difference is small. 
	
	\section{Conclusion and Future Work}
	\label{sec: concl}
	In this paper we have proposed a reduced complexity, content-customised, solution that can predict the bitrate ladder for adaptive streaming, based on spatio-temporal features extracted from uncompressed video at its native resolution. Our method predicts the intersection points of the RQ curves across spatial resolutions with a small number of video encodings and then parameterises a set of equations that predict the PF RQ points at the target bitrate ladder rungs. This enables construction of the bitrate ladder via constrained sampling of the quality and bitrate set of values. The proposed method was compared against two benchmarks, an exhaustive-search method (which produces the most accurate PF) and a more conventional interpolation-based method. When compared to the exhaustive search, the results show a mean BD-Rate loss of only 1.78\% and a mean BD-PSNR of 0.04 dB, but with a reduction on average of 89.06\% in the number of encodings needed. Although the BD statistics of the interpolation based method are better than those of the feature-based method, the latter provides a significant reduction of encodes, 61.46\% on average, to build the bitrate ladder. A hybrid method as a combination of the feature-based and the interpolation-based is examined resulting to a 1.26\% mean BD-Rate for only an additional 32.32\% number of encodings on average. Both FL and HL result in bitrate ladders that are composed on average over 80\% by Pareto optimal points. Adopting FL or HL could result in significant savings in processing time and energy consumption.
	
	Future work will focus on testing the effectiveness of the method across codecs and by employing different quality metrics. Firstly, as explained, the proposed method was developed for and tested on an HEVC codec. However, if the regression models were trained with data derived from a different codec, then we expect the performance gains to be comparable. Additional gains may however be possible by exploiting the correlation between content features and rate-distortion characteristics across different codecs. This could lead to even higher efficiency in estimating the bitrate ladder and in the overall content delivery. Furthermore, as video providers are using other quality metrics, e.g. VMAF or SSIM, to construct the bitrate ladders, we will test the proposed method on an extended set of quality metrics.

	\section*{Acknowledgements}
	\label{sec: ack}
	The authors would like to thank Dr Mariana Afonso and Kyle Swanson from Netflix Video Coding Group for all the insightful discussions that helped improving this work.

	\bibliographystyle{IEEEbib}
	\bibliography{refs1}

\end{document}